\documentclass[aip,pop,reprint,amsmath,amssymb,superscriptaddress]{revtex4-1}

\usepackage{graphicx}
\usepackage{bm}
\renewcommand{\vec}[1]{\boldsymbol{#1}}

\newcommand{\grad}{\nabla}
\renewcommand{\div}{\nabla \cdot}
\newcommand{\rot}{\nabla \times}

\def\grl{{Geophys. Res. Lett.} }
\def\jgr{{J. Geophys. Res.} }

\def\pop{{Phys. Plasmas} }

\def\prl{{Phys. Rev. Lett.} }
\def\ssr{{Space Sci. Rev.} }

\begin{document}

\title{Kinetic aspects of the ion current layer in a reconnection outflow exhaust}
\date{Received 12 May 2013; accepted 29 August 2013; published online 25 September 2013}

\author{Seiji Zenitani}
\affiliation{National Astronomical Observatory of Japan, 2-21-1 Osawa, Mitaka, Tokyo 181-8588, Japan.}
\email{seiji.zenitani@nao.ac.jp}
\author{Iku Shinohara}
\affiliation{Institute of Space and Astronautical Science, Japan Aerospace Exploration Agency, 3-1-1 Yoshinodai, Chuo, Sagamihara, Kanagawa 252-5210, Japan}
\author{Tsugunobu Nagai}
\affiliation{Tokyo Institute of Technology, Tokyo 152-8551, Japan}
\author{Tomohide Wada}
\affiliation{National Astronomical Observatory of Japan, 2-21-1 Osawa, Mitaka, Tokyo 181-8588, Japan.}

\begin{abstract}
Kinetic aspects of the ion current layer
at the center of a reconnection outflow exhaust near the X-type region
are investigated by a two-dimensional particle-in-cell (PIC) simulation. 
The layer consists of magnetized electrons and unmagnetized ions that carry a perpendicular electric current.
The ion fluid appears to be nonideal, sub-Alfv\'{e}nic, and nondissipative.
The ion velocity distribution functions contain multiple populations,
such as global Speiser ions, local Speiser ions, and trapped ions.
The particle motion of the local Speiser ions
in an appropriately rotated coordinate system
explains the ion fluid properties very well.
The trapped ions are the first demonstration of the regular orbits
in the chaotic particle dynamics [Chen and Palmadesso, {J. Geophys. Res.}, {\bf 91}, 1499 (1986)]
in self-consistent PIC simulations. 
They would be observational signatures in the ion current layer near reconnection sites.
\copyright~
2013 Author(s). All article content,
except where otherwise noted, is licensed under a Creative Commons Attribution 3.0 Unported License.
[{\tt http://dx.doi.org/10.1063/1.4821963}]
\end{abstract}


\maketitle

\section{Introduction}

Over the past few decades,
there has been steady progress in understanding
the physics of magnetic reconnection.
It is widely known that
the reconnection process is a complex multi-scale process,
in which small-scale physics in the X-type region critically
controls the large-scale system evolution.
In the magnetohydrodynamics (MHD),
such a small-scale physics is often represented by
a spatial or functional profile of an electric resistivity in the Ohm's law.\citep{ugai92}
In a collisionless plasma, such as in the Earth's magnetosphere,
plasma kinetic motion plays a role as an effective resistivity.
However, particle motion is highly sensitive to the local electromagnetic structure.
Therefore, kinetic reconnection researches have focused on
the structure and the relevant kinetic physics
in and around the X-type region. 
Particular attention has been paid to
the deviation from the ideal Ohm's law or the ideal condition
$\vec{E}+\vec{V}_s\times\vec{B} = 0$
for each species $s$,
because the violation of the ideal condition 
is necessary to transport
magnetic flux across the X-line.

Earlier expectations agree that
the X-type region has a two-scale structure,\citep{hesse01,drake07}
an outer layer
in which ions decouple from the magnetic field
and
an inner layer in which electrons are unmagnetized.
Inside the outer layer but outside the inner layer,
the relative motion between unmagnetized ions and magnetized electrons
gives rise to Hall effects. 
The Hall effects introduce characteristic signatures
to the reconnection site.
For example, \citet{sonnerup79} predicted that
in-plane current loops generate quadrupolar magnetic field perturbations.
This is interpreted as the three-dimensional modulation of
the magnetic topology.\citep{terasawa83,mandt94}
These signatures have been verified
by kinetic particle-in-cell (PIC) simulations,\citep{prit01,hesse01}
by {\it in-situ} observation
in the Earth's magnetosphere,\citep{nagai01,oieroset01,mozer02,runov03}
and
by laboratory experiments.\citep{ren05,whm05}

The structure of the inner electron layer has been revealed
by modern PIC simulations.\citep{dau06,kari07,shay07,hesse08}
The layer consists of a central dissipation region\citep{zeni11c} and
bi-directional electron jets.\cite{kari07,shay07,hesse08} 
The jet was initially suspected to be an outer part of the dissipation region,
but recent works suggested that it is non-dissipative.\citep{hesse08,zeni11d} 
Signatures of the electron jet have been reported
in the magnetosheath\citep{phan07} and
in laboratory experiments.\citep{ren08}
More recently, the Geotail spacecraft observed
both the central dissipation region and the bi-directional electron jets
in magnetotail reconnection.\citep{nagai11,zeni12} 
Therefore, despite several uncertainties,
the structure of the electron layer is fairly well understood.

Meanwhile,
much less is known about the reconnection outflow region. 
Early investigations were conducted by hybrid simulations,
originally motivated by the potential slow-shock formation
at the separatrices.\citep{kv95,lin96,higashimori12}
The following works have revealed
the internal structure of an outflow exhaust\cite{lot98,arz01} and
associated ion dynamics.\citep{nakabaya97,nakamura98,lot98,arz01,aunai11a}
For example, there is usually an current layer at the center of the exhaust.\citep{lot98,arz01,higashimori12}
However, since hybrid models ignore electron physics,
it is not clear whether these results are reliable near the X-type region.
Recent large-scale PIC simulations focused on
basic processes in an outflow exhaust beyond the X-type region. 
\citet{drake09} studied pickup-type ion heating
at the boundary layer of the outflow exhaust.
\citet{liu11a,liu12} investigated the role of
the pressure anisotropy in the lateral evolution of the exhaust. 
At present, it is not clear
how the three domains are connected:
the X-type region, the electron nonideal layer,
and the outflow exhaust.

In this work,
by means of PIC simulations,
we investigate kinetic aspects of an ion current layer
at the center of the outflow exhaust
just downstream of the electron nonideal layer.
In Sec.~\ref{sec:setup},
we briefly describe our numerical setup. 
In Sec.~\ref{sec:results},
we overview our simulation results, and then
we study ion fluid properties that could deviate from the ideal MHD
in the ion current layer. 
Next we examine ion velocity distribution functions (VDFs) and
the relevant particle motions.
Impacts to the ion fluid
properties are also discussed. 
In Sec.~\ref{sec:discussion},
we discuss some generic issues,
such as the magnetic diffusion and the ion outflow speed. 
Section \ref{sec:summary} contains a summary.
In the Appendix section,
we describe supplemental test-particle simulations
to better understand single-particle dynamics in the current layer.

\begin{figure*}[tbp]
\centering
\includegraphics[width={2\columnwidth},clip]{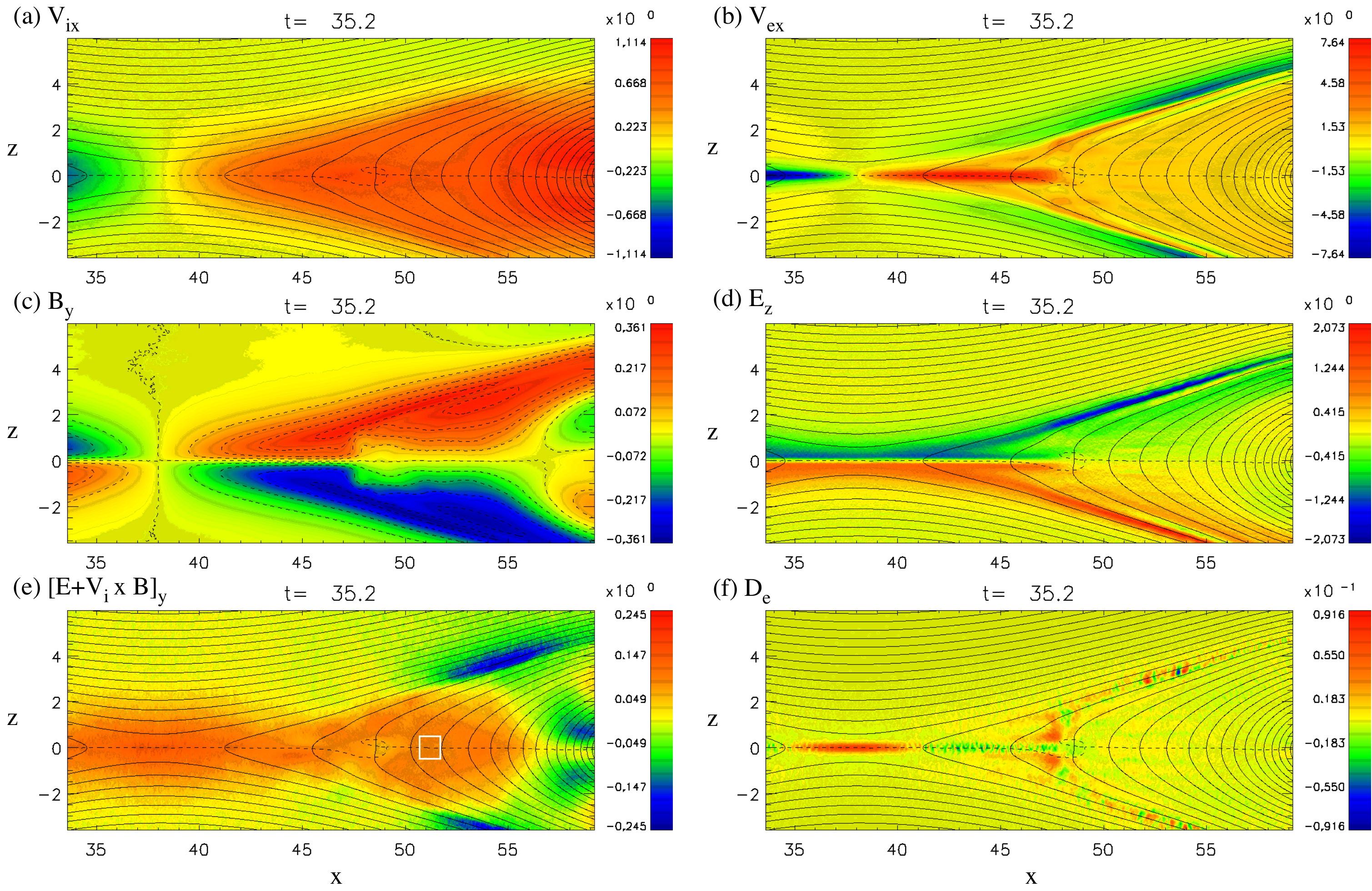}
\caption{(Color online)
Results of the main run, averaged over $t=35$--$35.25$.
The contour lines are in-plane magnetic field lines and
the dashed line indicates the field reversal, $B_x=0$.
(a) The ion outflow speed $V_{ix}$ in a unit of $c_{Ai}$,
(b) the electron outflow speed $V_{ex}$,
(c) the out-of-plane magnetic field $B_y$ and its contour,
(d) the vertical electric field $E_z$ in a unit of $c_{Ai}B_0$, and
(e) the out-of-plane component of the ion Ohm's law, $[\vec{E}+\vec{V}_{i}\times\vec{B}]_y$ in a unit of $c_{Ai}B_0$. The white box is the region of our interest in Sec.~\ref{sec:dist}.
(f) The nonideal dissipation measure $D_e$ in a unit of $j_0c_{Ai}B_0$.
}
\label{fig:outflow}
\end{figure*}

\begin{figure}[tbp]
\centering
\includegraphics[width={\columnwidth},clip]{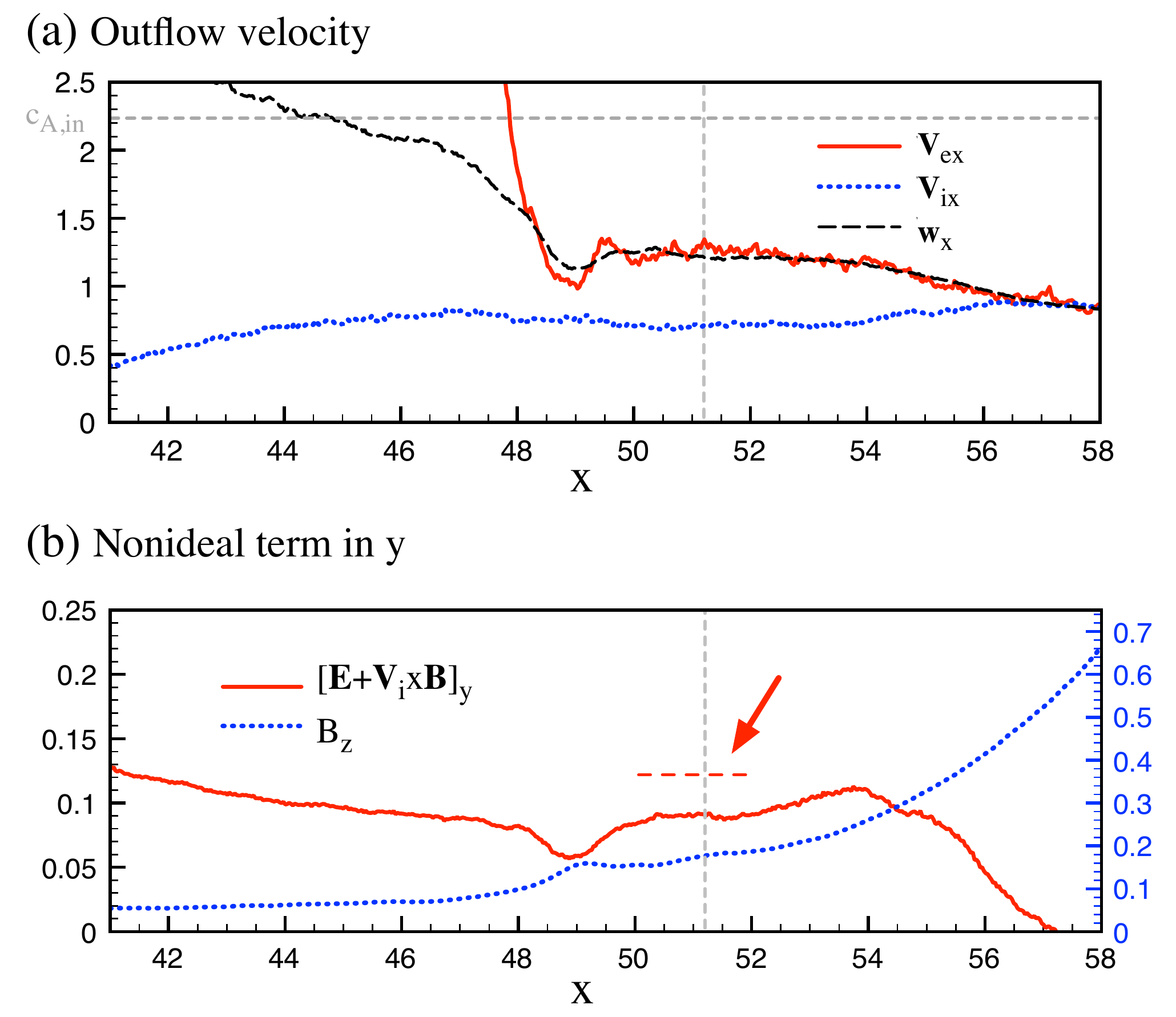}
\caption{(Color online)
(a) Plasma outflow velocities $V_{ix}$ and $V_{ex}$, and the \textbf{E} $\times$ \textbf{B} speed $w_{x}$ at $z=0$.
The dashed horizontal line indicates the initial inflow Alfv\'{e}n speed, $c_{A,in}=2.24$.
(b) The nonideal term in ion Ohm's law $[\vec{E}+\vec{V}_i\times\vec{B}]_y$ and
the magnetic field $B_z$ at $z=0$.
The dashed vertical lines indicate $x=51.2$.
}
\label{fig:horizontal}
\end{figure}

\begin{figure}[htbp]
\centering
\includegraphics[width={\columnwidth},clip]{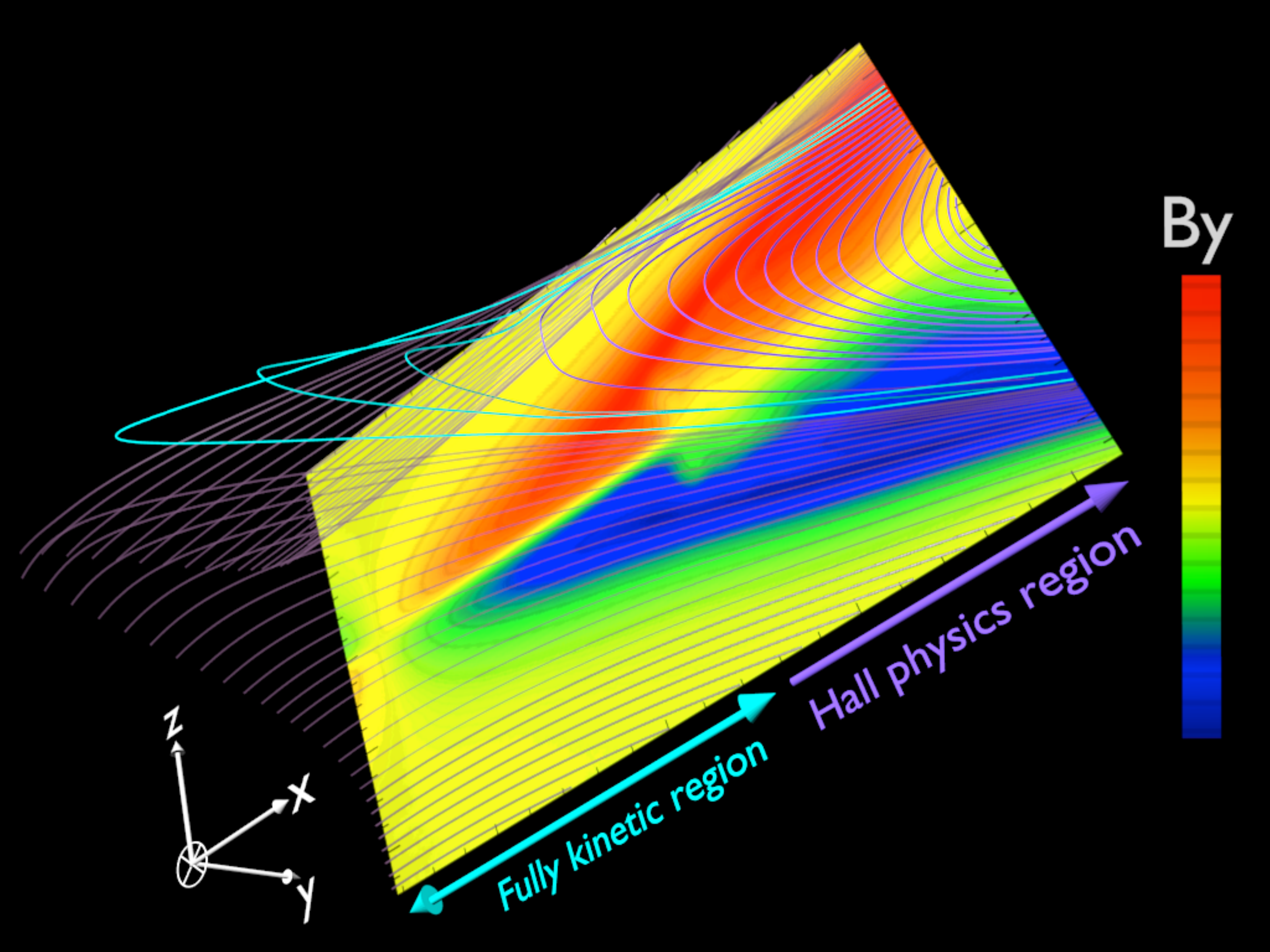}
\caption{
A 3D view of the magnetic field lines.
The rear panel (color contour) indicates the out-of-plane Hall field, $B_y$.
The blue field lines cross the midplane inside the electron nonideal layer (fully kinetic region).
\label{fig:3D}
}
\end{figure}

\section{PIC simulation}
\label{sec:setup}

We carry out simulations with
a partially-implicit PIC code.\citep{hesse99}
The results are presented in normalized units: 
lengths to the ion inertial length $d_i=c/\omega_{pi}$,
times to the inverse ion cyclotron frequency $\Omega_{ci}^{-1}=m_i/(eB_0)$, and
velocities to the typical ion Alfv\'{e}n speed $c_{Ai}=B_0/(\mu_0 m_i n_0)^{1/2}$. 
Here, $\omega_{pi}=(e^2n_0/\varepsilon_0m_i)^{1/2}$ is the ion plasma frequency,
$n_0$ is the reference plasma density, and
$B_0$ is the background magnetic field.
We employ a Harris-like configuration,
$\vec{B}(z)=B_0 \tanh(z/L) \vec{e}_x$ and
$n(z) = n_{0} [0.2 + \cosh^{-2}(z/L)]$,
where $L=0.5 d_i$ is the half thickness of the current sheet.
Ions are assumed to be protons.
The ion-electron mass ratio is $m_{i}/m_{e}=100$,
the temperature ratio is $T_e/T_i=0.2$, and
the frequency parameter is $\omega_{pe}/\Omega_{ce}=4$,
where $\omega_{pe}$ is the electron plasma frequency and
$\Omega_{ce}$ is the electron gyro frequency.
The computational domain is $x, z \in[0,76.8]\times[-19.2, 19.2]$. 
It is resolved by $2400 \times 1600$ grid cells.
Periodic ($x$) and reflecting ($z$) boundaries are used. 
In order to trigger reconnection,
a small initial perturbation is introduced into current density and magnetic fields.
The typical amplitude of the perturbation magnetic field is $\delta B=0.1 B_0$.
It is localized at the center of the simulation domain.
The run is almost the same as run 1A in our previous work\citep{zeni11d}
except that we have halved a timestep to
$\Omega_{ci}\Delta t=0.000625$ ($\omega_{pe}\Delta t=0.25$) for safety. 
We use $1.7 \times 10^9$ particles ($\approx 200$ pairs in an upstream cell).

\section{Results}
\label{sec:results}

\subsection{Overview}
\label{sec:overview}

The reconnection process starts around
the center of the simulation domain. 
The evolution is virtually identical to
that of Run 1A in our previous work.\citep{zeni11d}
The system evolves in a time scale of 10 $[\Omega_{ci}^{-1}]$. 
Panels in Figure \ref{fig:outflow} show
selected quantities at $t=35$.
They are averaged over a time interval of $0.25$ to remove noises.
The X-line is near the center of the simulation domain, $(x,z)\sim(38.1,0)$.
At this time,
the reconnection process goes on at a quasisteady rate and
the structure of the reconnection site is well-developed.

Figure \ref{fig:outflow}(a) shows
the ion outflow speed $V_{ix}$. 
Bi-directional ion jets emanate from the X-line and then 
the ion outflow speed increases with $x$.
Figure \ref{fig:outflow}(b) shows
the electron outflow speed $V_{ex}$.
A narrow electron jet also emanates from the X-line,
but it is embedded inside the large-scale ion flow. 
The jet is popularly denoted to as the super-Alfv\'{e}nic electron jet,\citep{kari07,shay07}
because its speed is considerably faster than the ion outflow speed,
$V_{ex}\gg V_{ix} \sim \mathcal{O}(c_{Ai})$.
In this case, the jet terminates at $x\approx 48$.
In addition,
there are field-aligned electron flows to the X-line along the separatrices,
as seen in light green in Figure \ref{fig:outflow}(b).
Figure \ref{fig:horizontal}(a) shows the 1D profiles of plasma outflows at the midplane $z=0$.
The \textbf{E} $\times$ \textbf{B} velocity is also presented,
$\vec{w}=(\vec{E}\times\vec{B})/B^2$.
The ion speed reaches its maximum $V_{ix}\approx 0.8$ around $x \approx 47$,
but it is substantially slower than $w_{x}$ and $V_{ex}$.
In the downstream of the super-Alfv\'{e}nic jet,
the electron speed drops to $w_{x}$ but
it remains faster than the ion speed, $V_{ex}>V_{ix}$.

Figure \ref{fig:outflow}(c) presents the out-of-plane magnetic field, $B_y$.
It exhibits a well-known quadrupole pattern
due to the Hall effect.\citep{sonnerup79}
The peak amplitude is $|B_y| \sim 0.36$.
This is consistent with recent PIC simulations by other groups.\citep{prit01b,drake09}
The outrunning electrons ($V_{ex}>V_{ix}$) and the field-aligned incoming electrons
are responsible for the in-plane current circuit to generate the quadrupole field $B_y$. 
We notice that $B_y$ changes its sign in the downstream region ($55<x$).
This is a Hall effect of another kind, first reported by {\citet{nakabaya97}.
The plasma outflow piles up the reconnected magnetic field ($B_z$) there.
Then the pileup field hits the preexisting plasma sheet in the farther downstream.
Due to their inertia, ions penetrate into the pileup region deeper than electrons in the $x$ direction.
The relevant in-plane current generates
an out-of-plane magnetic field $B_y$ of the opposite polarity.

Figure \ref{fig:outflow}(d) presents the vertical electric field $E_z$.
It is distinctly strong along the separatrices. 
This is basically the polarization electric field,
directed from the low-density side to the high-density side.
It is often referred to as the Hall electric field
around the X-line.\citep{shay98,arz01,prit01b,wygant05} 
Inside the outflow exhaust,
$E_z$ is negative in the upper half and positive in the lower half. 
This corresponds to the $+x$-ward transport of the Hall magnetic field $B_y$.
In the downstream region ($55<x$) where $B_y$ changes the sign,
$E_z$ accordingly changes the sign.

Figure \ref{fig:outflow}(e) shows
the out-of-plane component of the ion Ohm's law,
$[\vec{E}+\vec{V}_i\times\vec{B}]_y$.
Since $E_y$ is responsible for the flux transport of in-plane magnetic fields,
it is sometimes used as an ion-scale proxy of the reconnection site. 
In fact, the nonideal region $[\vec{E}+\vec{V}_i\times\vec{B}]_y \ne 0$ is
wide spread over the X-type region.
Figure \ref{fig:horizontal}(b) shows its 1D profile along the outflow line.
The reconnected magnetic field $B_z$ is also presented.

Figure \ref{fig:outflow}(f) shows
an energy dissipation measure,\citep{zeni11c}
\begin{eqnarray}
\label{eq:De}
D_e =
\gamma_e \big[ \vec{j} \cdot ( \vec{E} + \vec{V}_e \times \vec{B} ) -
\rho_c ( \vec{V}_e \cdot \vec{E} ) \big],
\end{eqnarray}
where $\gamma_e=\sqrt{1-(V_e/c)^2}$ is the Lorentz factor and $\rho_c$ is the charge density.
This can be reduced to
$D_e \approx \vec{j} \cdot ( \vec{E} + \vec{V}_e \times \vec{B} ) \approx \vec{j} \cdot ( \vec{E} + \vec{V}_i \times \vec{B} )$
in a nonrelativistic quasi-neutral plasma.
This stands for the energy transfer from the electromagnetic field to the plasma
in the electron's rest frame, and
corresponds to the nonideal energy dissipation. 
An energy dissipation region is located near the X-line, $35 \lesssim x \lesssim 41$. 
It is compact and is deeply embedded
inside the wider ion nonideal region (Fig.~\ref{fig:outflow}(e)).
There are other dissipative structures in Figure \ref{fig:outflow}(f).
The small dissipative and anti-dissipative spots along the separatrices
will be due to electrostatic-type instabilities.\citep{cattell05,keizo06d}

As already stated,
the super-Alfv\'{e}nic electron jet terminates at $x \approx 48$.
Farther downstream ($x \gtrsim 48$),
the electron speed quickly decreases to $V_{ex} \approx w_{x}$ (Fig.~\ref{fig:horizontal}(a)). 
Instead, $B_z$ increases at $x \approx 49$.
Electrons are unmagnetized in the upstream jet, while
they are magnetized by the compressed magnetic field $B_z$ in the downstream.
On the other hand, ions are magnetized neither in the upstream nor in the downstream.
In this sense, this is a transition layer
between fully kinetic region and Hall-physics region. 
Reference~\onlinecite{zeni11d} called the layer the ``electron shock.''
It propagates in the downstream direction
as shown in Fig. 8(a) in Ref.~\onlinecite{zeni11d}.
Strictly speaking, it differs from standard MHD shocks, but
it is a shock-like jump structure
across which plasmas and the magnetic field are transported.
The vertical structures at $x\approx 47$--$48$ (in red; Fig.~\ref{fig:outflow}(f))
suggest secondary energy dissipation near the transition layer,
probably due to shock-driven electron flows.

Figure \ref{fig:3D} shows the three-dimensional structure of the magnetic field lines
with the rear panel of $B_y$.
To better understand the topology,
we manually set foot-points of the field lines.
Therefore the field-line density is not proportional to $|\vec{B}|$,
but it provides a good qualitative picture. 
As can be seen, the field lines are dragged out from the original $x$-$z$ plane
to the $-y$ direction.\citep{mandt94} 
The field lines between the X-line and the electron shock are in light blue color,
corresponding to the fully kinetic region at the midplane.
The field lines are twisted at the separatrices
between the inflow regions and the blue region. 
The field lines are again twisted
between the blue region and the outflow region in purple.
The drag-angle of the field lines changes here. 
This is associated with the electron shock.
At the shock, the electrons are trapped by the magnetic field lines and then
they travel in the field-aligned directions. 
Such electron flows can be seen
at $(x,z) \approx (48, \pm 1.5)$ in Figure \ref{fig:outflow}(b). 
The resulting electric current modifies the magnetic topology at this boundary. 
Careful inspection of Figure \ref{fig:outflow}(c) reveals
a double-peak structure in $B_y$. 
This corresponds to a two-scale structure of the magnetic field,
the blue region and the outflow region in Figure \ref{fig:3D}.
The field lines remain tilted in the downstream,
because they are still dragged to the $-y$ direction by the Hall effect.

We examine the structure downstream the electron shock. 
Here, electrons are magnetized and
their speed is similar to the \textbf{E} $\times$ \textbf{B} speed, 
$V_{ex} \approx {w}_{x} \approx 1.2$
(Fig.~\ref{fig:horizontal}(a)). 
The ion speed slightly decreases to
its typical value $V_{ix} \approx 0.7$ around $50<x<54$. 
MHD theories expect that
the reconnection outflow speed is approximated by the Alfv\'{e}n speed in the inflow region.
In this case, the inflow Alfv\'{e}n speed is initially $c_{A,in} = 2.24$.
This is indicated by the dashed horizontal line in Figure \ref{fig:horizontal}(a).
If normalized by upstream quantities at $|z| = 3$ near the X-line at $t=35$,
the inflow Alfv\'{e}n speed is $c_{A,in} \approx 1.62$.
The ion outflow speed is still slow, $V_{ix} < c_{A,in}$.

The relative motion of ions and electrons sustains
a Hall current $J_x$ in the outflow channel.\citep{sonnerup79}
In the shock-upstream ($x\lesssim48$), both electrons and ions are unmagnetized,
and so they can travel in the perpendicular direction to
the local magnetic field.
In particular, due to their light mass,
electrons are the main carrier of the electric current
for $B_x$- and $B_y$-reversals.
In contrast, in the shock-downstream ($48 \lesssim x$),
electrons are magnetized and
only ions can travel in the perpendicular direction.
So, in the \textbf{E} $\times$ \textbf{B} frame (deHoffmann--Teller frame),
unmagnetized ions carry the most of the perpendicular current
with respect to the local magnetic field $\vec{B}\approx B_z\vec{e}_z$. 
In this sense, ions are the main current carriers in the shock-downstream.
Therefore, we call the shock-downstream region the ``ion current layer.'' 

In order to sustain the field reversals
without the electron perpendicular flow,
the system needs a broader current layer
in the shock-downstream ($48 \lesssim x \lesssim 56$)
than in the shock-upstream. 
The broad current layer corresponds to
a broad cavity around the midplane ($z \approx 0$)
between the $B_y$ regions in Figure \ref{fig:outflow}(c). 
As a consequence, we can see a step-shaped pattern in the $B_y$-profile.
Similar patterns in $B_y$ can be seen
in recent simulations at sufficiently high mass-ratios.\citep{le10a,shay11}

Another macroscopic signature of the ion current layer is
the violation of the ion ideal condition. 
As shown in Figures \ref{fig:outflow}(e) and \ref{fig:horizontal}(b),
the ion Ohm's law remains nonzero.
At the midplane ($z=0$), the nonidealness arises from
the slow ion motion with respect to the \textbf{E} $\times$ \textbf{B} speed,
$[\vec{E}+\vec{V}_i\times\vec{B}]_y \approx (w_{x}-V_{ix}) B_z > 0$.
As indicated by the red arrow in Figure \ref{fig:horizontal}(b),
it is almost flat $\partial_x \approx 0$
downstream the shock ($50 \lesssim x \lesssim 52$). 
This is more evident in an earlier stage $t=30$,
when there was more room in the shock downstream. 
Despite the limited system size in $x$,
the flat cavity structure (Fig.~\ref{fig:outflow}(c)),
the magnetic angle (Fig.~\ref{fig:3D}), and
the velocity profile (Fig.~\ref{fig:horizontal}(a)) indicate that
the structure of the ion current layer is approximately invariant in $x$. 

\begin{figure}[tbp]
\centering
\includegraphics[width={\columnwidth},clip]{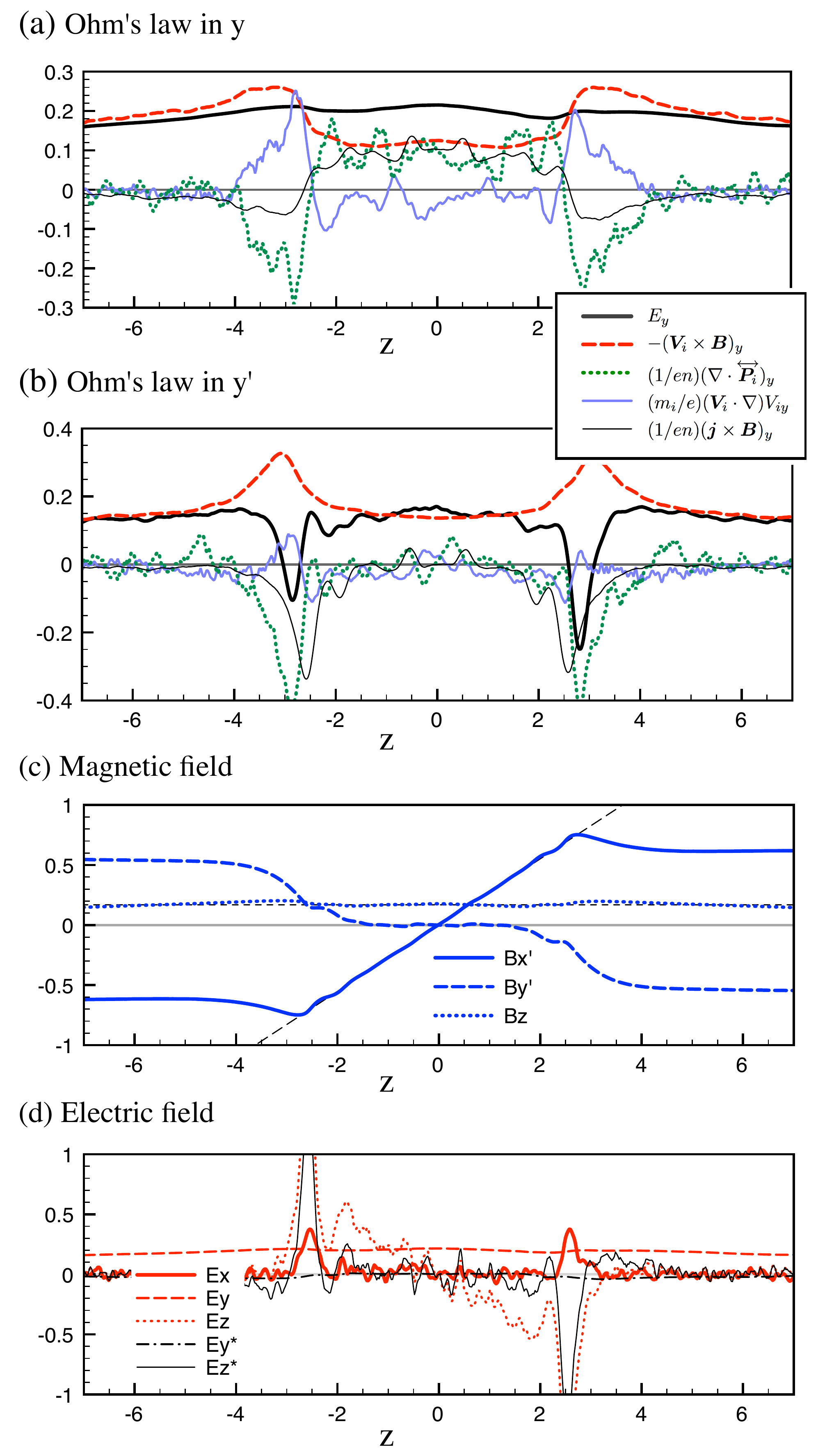}
\caption{(Color online)
The composition of the ion Ohm's law (See Eqs.~\ref{eq:iohm} and \ref{eq:gohm}) at $x=51.2$.
(a) The $y$ component and (b) the $y'$ component in the rotated coordinate system.
(c) Magnetic field in the rotated coordinate system.
(d) Electric field $\vec{E}$ in the simulation frame and $\vec{E}^*$ in the moving frame.
}
\label{fig:vertical}
\end{figure}

\subsection{Vertical structure}
\label{sec:vertical}

Next we study several quantities
in a vertical cut of the ion current layer. 
We first examine the ion Ohm's law.
It can be decomposed in the following ways,
\begin{eqnarray}
\label{eq:iohm}
\vec{E} + {\vec{V}_i}\times\vec{B}
&=&
\frac{1}{en} \div \overleftrightarrow{\vec{P}_i}
+ \frac{m_i}{e} \Big( (\vec{V}_i\cdot\grad)\vec{V}_i+\frac{\partial \vec{V}_i}{\partial t}\Big)
~~~\\
\label{eq:tohm}
&=&
\frac{1}{en} \div \overleftrightarrow{\vec{T}_i}
+ \frac{m_i}{e} \frac{\partial \vec{V}_i}{\partial t}
\\
\label{eq:gohm}
&=&
\frac{1}{en} \vec{j}\times\vec{B}
- \frac{1}{en} \div \overleftrightarrow{\vec{P}_e}
- \frac{m_e}{e} \frac{d \vec{V}_e}{d t}
.
\end{eqnarray}
Here, $\overleftrightarrow{\vec{P}_s}$ is the pressure tensor and
$\overleftrightarrow{\vec{T}_s}$ is the momentum flux density tensor,
$T_{jk} = m\int f v_{j}v_{k} d\vec{v}$.
In Eq.~\ref{eq:iohm},
the terms in the right hand side are
the divergence of the ion pressure tensor and
the ion bulk inertial term.
The second form (Eq.~\ref{eq:tohm}) emphasizes
the physical meaning of the momentum transport.
The third form (Eq.~\ref{eq:gohm}) is identical to
the so-called generalized Ohm's law in the $(m_e/m_i) \ll 1$ limit. 

In Figure \ref{fig:vertical}(a),
we show the $y$ component of the composition of the ion Ohm's law
at $x=51.2$.
This position is indicated by the dashed vertical lines
in Figures \ref{fig:horizontal}(a) and \ref{fig:horizontal}(b).
Some quantities look noisy,
in particular, the pressure tensor term,
because the time interval $0.25$ is
shorter than typical ion gyroperiod $2\pi$.
The sum of the pressure tensor term (green dotted line) and
the bulk inertial term (blue line) 
is usually equal to
the Hall term $\vec{j}\times\vec{B}$ in the generalized form (Eq.~\ref{eq:gohm}).
Outside the central current layer, 
the pressure tensor term and the bulk inertial term
tend to cancel each other. 
This indicates that
the outer layers play a minor role in the $y$-momentum balance,
$[\div \overleftrightarrow{\vec{T}_{i}}]_y \approx 0$.

As shown in Figures \ref{fig:3D},
the magnetic field lines are dragged out from
the initial $x$-$z$ plane due to the Hall effect. 
To better understand the magnetic topology,
we consider
an appropriately rotated coordinate system.\citep{hesse08,klimas12}
The magnetic field transforms like
\begin{eqnarray}
\left\{
\begin{array}{l}
B_{x'} = B_x \cos \alpha + B_y \sin \alpha \\
B_{y'} = -B_x \sin \alpha + B_y \cos \alpha
\end{array}
\right.
\end{eqnarray}
where $\alpha$ is a rotation angle, and then
we assume that the magnetic field lies in the $x'$-$z$ plane.
At $x=51.2$ of our interest, we obtain $\alpha \approx 41^{\circ}$
by minimizing $\sum B_{y'}^2$ in $-1.4<z<1.4$. 
Figure \ref{fig:vertical}(c) shows the magnetic field in the rotated coordinate system. 
The magnetic field is approximated by a parabolic field,
\begin{eqnarray}
\label{eq:B-fit}
B_{x'}=0.276z,~~B_{y'}=0,~~B_z=0.17,
\end{eqnarray}
as indicated by dashed lines in Figure \ref{fig:vertical}(c). 
Small plateaus at $2<|z|<2.5$ corresponds to the electron jet region
that separates the blue region and the outflow region in Figure \ref{fig:3D}.
Figure \ref{fig:vertical}(b) shows
the composition of the ion Ohm's law in the $y'$ direction at $x=51.2$.
In this coordinate system,
the electric field $E_{y'}$ (black line)
is balanced by the convection electric field (red dashed line).
This tells us that
the ion motion is nearly ideal in the $x'$-$z$ plane.\citep{hesse08}
The ions comove with the magnetic field,
in the sense that they follows the \textbf{E} $\times$ \textbf{B} drift
in the rotated plane.
In contrast, 
the electric field is nonideal in $x'$ (not shown).
The nonideal part is balanced by the momentum transport term,
$[\vec{E}+\vec{V}_i \times \vec{B}]_{x'}
\approx (1/en) \partial_{x'} P_{ix'z}
\approx  (1/en) \partial_{x'} T_{ix'z}$. 
The electric current $\vec{j}$ is primarily in the $y'$ direction,
consistent with the magnetic field reversal in $x'$. 
The minimization of $\sum j_{x'}^2$ gives a similar angle $\alpha \approx 44^{\circ}$. 
Consistent with Figure \ref{fig:outflow}(f),
the nonideal energy dissipation is negligible,
because $D_e \approx j_{y'} [\vec{E}+\vec{V}_i\times\vec{B}]_{y'} \approx 0$.

Shown in Figure \ref{fig:vertical}(d) are the electric fields
across the ion current layer at $x=51.2$.
The $E_x$ is noisy but small. 
The reconnection electric field $E_y$ is fairly constant (see also Fig.~\ref{fig:vertical}(a))
and $E_z$ looks bipolar (see also Fig.~\ref{fig:outflow}(d))
except for the separatrices. 
They are basically the motional electric fields of
the reconnected field $B_z$ and the Hall field $B_y$.\citep{arz01} 
Keeping this in mind,
we consider an appropriately moving frame at the velocity of $\vec{U}$.
The electric field $\vec{E}^*$ in the moving frame is
$\vec{E}^*=\vec{E}+\vec{U}\times\vec{B}$.
By minimizing $\sum |E^{*2}|$ over $x, z \in [50.7,51.7] \times [-0.5,0.5]$,
we obtain $\vec{U}=(1.2,0,0)$.
The electric fields $\vec{E}^*$
are overplotted in Figure \ref{fig:vertical}(d).
The $x$ component is unchanged, $E_x^*=E_x$.
Except for the separatrices ($z \sim \pm 2.5$),
all three components of $\vec{E}^*$ are fairly small,
as reported by a previous work.\citep{drake09}
When $(c^2\vec{B}^2-\vec{E}^2) \ge 0$ and $\vec{E} \cdot \vec{B} = 0$,
it is possible to transform into the frame,
in which the electric field vanishes. 
These conditions are fairly satisfied here.
Note that $\vec{U}$ does not always equal to
the local \textbf{E} $\times$ \textbf{B} velocity $\vec{w}$. 
Indeed, $\vec{U}=\vec{w}$ at the midplane but
$\vec{U}\ne\vec{w}$ away from the midplane. 
It is important that
a single nonlocal velocity $\vec{U}$ transforms away the electric field. 
This suggest that the magnetic structure travels with $\vec{U}$. 

Judging from these results,
the magnetic structure of the ion current layer
will be best understood in
the rotated coordinate system in the moving frame with $\vec{U}$.
Hereafter we refer to this rotated, $\vec{U}$-moving frame as the ``reference frame.''

\begin{figure*}[ht]
\centering
\includegraphics[width={2\columnwidth},clip]{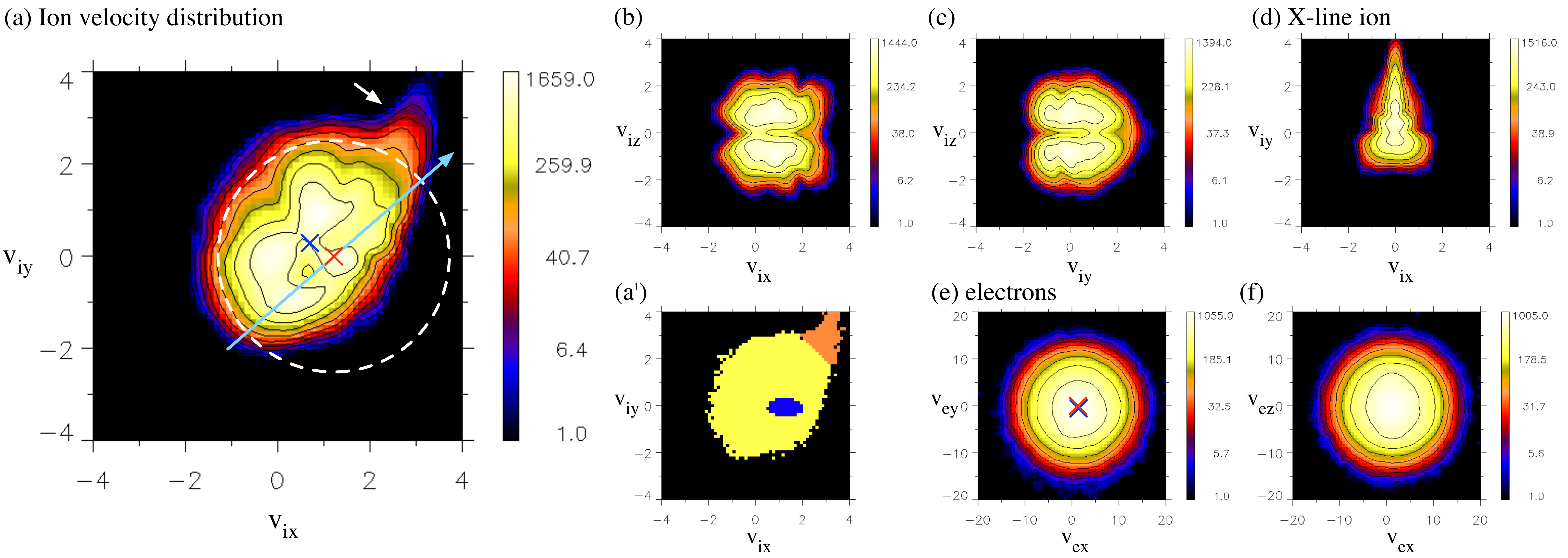}
\caption{(Color online)
Ion velocity distribution functions in the central current layer,
integrated over $x,z \in [50.7,51.7] \times [-0.5,0.5]$ at $t=35$:
(a) $\log N_i(v_x,v_y)$, (b) $\log N_i(v_x,v_z)$, and (c) $\log N_i(v_y,v_z)$.
(d) Ion velocity distribution function $\log N_i(v_x,v_y)$ near the X-line
($x,z \in [37.6,38.6] \times [-0.5,0.5]$).
Electron velocity distribution functions in the ion current layer
($x,z \in [50.7,51.7] \times [-0.5,0.5]$):
(e) $\log N_e(v_x,v_y)$ and (f) $\log N_e(v_x,v_z)$.
In Panels (a) and (e), the red and blue crosses indicates
the reference-frame velocity $\vec{U}$ and
the average ion (electron) velocity $\vec{\bar{V}}_i$ ($\vec{\bar{V}}_e$).
}
\label{fig:outPDF}
\end{figure*}

\subsection{Distribution function}
\label{sec:dist}

To gain further insights into the ion kinetic physics,
we examine the plasma VDF and the relevant particle trajectories.
Figures \ref{fig:outPDF}(a)--\ref{fig:outPDF}(c) show the ion VDFs
integrated over $x, z \in [50.7,51.7] \times [-0.5,0.5]$ at $t=35$.
This volume is indicated by the white box in Figure \ref{fig:outflow}(e).
The VDFs consist of $3.9 \times 10^5$ ions.
Particles are integrated in the third directions in each panel. 
The VDFs are box-car averaged over the neighboring grid points to better see the structure. 
Figure \ref{fig:outPDF}(a) presents the ion VDF in $v_x$-$v_y$.
The blue and red crosses indicate
the average ion velocity $\vec{\bar{V}}_i = (0.69,0.29,0)$ in this volume,
and the reference-frame velocity $\vec{U}$, respectively.
The blue arrow indicates the oblique direction
(the ${x'}$ direction; rotated by $\alpha = 41^{\circ}$),
discussed in Sec.~\ref{sec:vertical}. 
Figures \ref{fig:outPDF}(b) and \ref{fig:outPDF}(c) are the ion VDFs in the other two velocity spaces.
The ions are split to two populations in the upper and lower halves in $z$,
because they are bouncing in $z$ in the ion current layer.
Similar VDFs were found around the midplane
in the outflow exhaust in previous works.\citep{arz01,drake09,aunai11a}
The bounce motion is evident
in the $v_{z}$-$z$ phase-space diagram around $x=51.2$ in Figure \ref{fig:phase}(a).
The counter-stream motion in ${\pm}z$ is dominant in $|z|<0.5$. 
Such counter $z$-motion is more significant in the VDFs in the shock-upstream,
because of the bipolar Hall electric field $E_z$.\citep{shay98,arz01,prit01b,wygant05} 
Figure \ref{fig:phase}(b) is another ion phase-space diagram in $v_{y'}$-$z$.
To see the ion motion in the reference frame,
we consider a relative velocity from $\vec{U}$.
It is in the $y'$ direction in the rotated coordinates. 

Panels in Figure \ref{fig:dist} show
the spatial distributions of the ions in the above VDF.
The red boxes indicate their location
($x, z \in [50.7,51.7] \times [-0.5,0.5]$) at $t=35$. 
The color indicates their position
in the $v_x$-$v_y$ space at $t=35$ in Figure \ref{fig:outPDF}(a'). 
Figure \ref{fig:traj} shows reconstructed ion orbits. 
We select representative ions in the above VDF (Figs.~\ref{fig:outPDF}(a)--~\ref{fig:outPDF}(c)) at $t=35$,
and then we track their orbits during $15<t<40$,
by test-particle simulations
in the electromagnetic fields of the PIC simulation. 
The PIC field data is sampled every 0.5 $[\Omega_{ci}^{-1}]$. 
Since the ion motion is insensitive to small fluctuations,
the reconstructed orbits agree with the PIC data very well. 
An error in the position is within 0.2 $[d_i]$ during $\Delta t=5$.
The orbits are presented in the $x$-$y$-$z$ space (Fig.~\ref{fig:traj}(a)),
in the $v_{x}$-$v_{y}$ space (Fig.~\ref{fig:traj}(b)), and
in the $v_{y'}$-$z$ space (Fig.~\ref{fig:traj}(c)). 
The rotation angle and the reference-frame velocity are fixed to
$\alpha=41^{\circ}$ and $\vec{U}=(1.2,0,0)$.
The $y$ position is set to $y=0$ at $t=35$. 
The circle marks the position at $t=35$ in Figures \ref{fig:traj}(b) and \ref{fig:traj}(c).

\begin{figure}
\centering
\includegraphics[width={\columnwidth},clip]{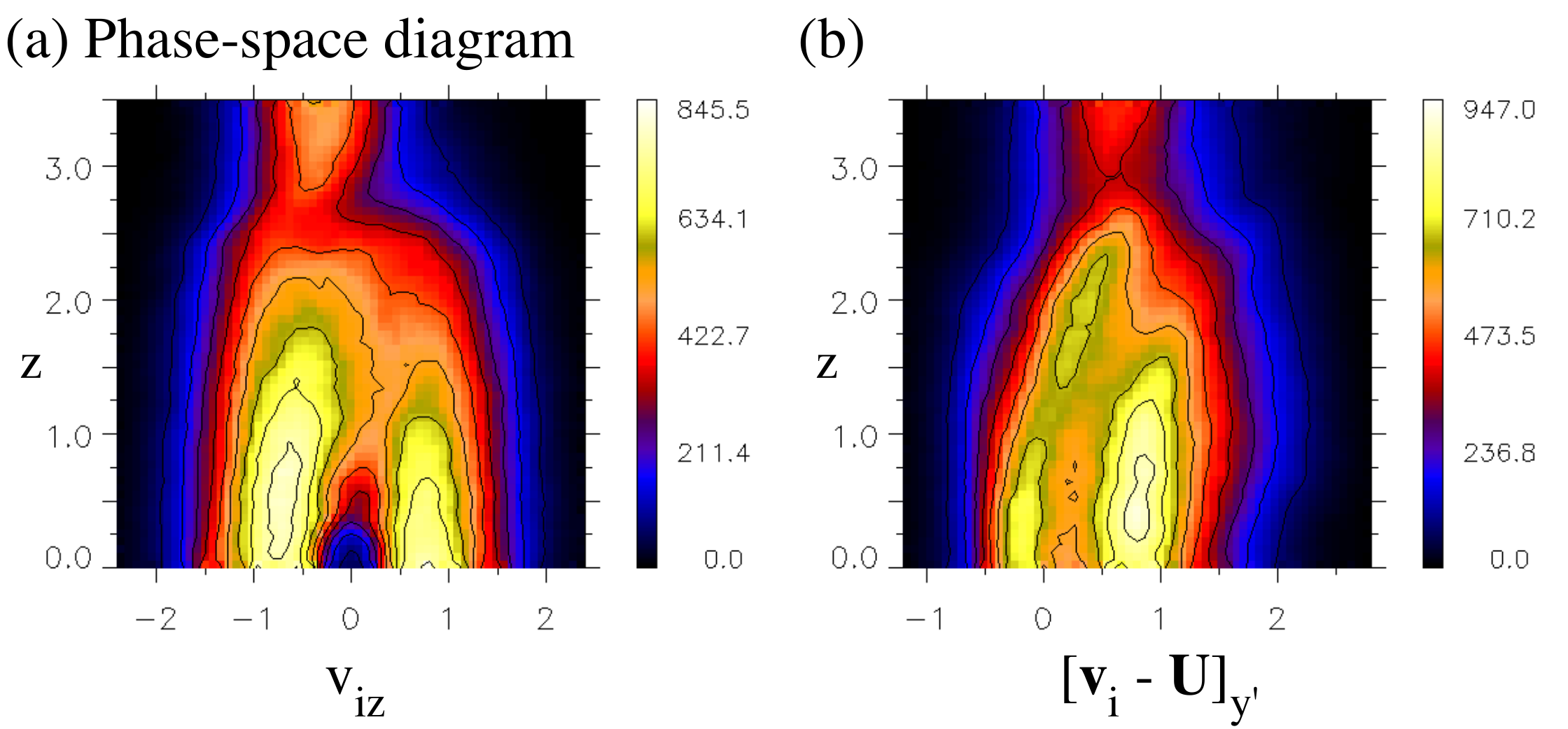}
\caption{(Color online)
Ion phase-space diagrams integrated over $x \in [50.7,51.7]$ at $t=35$:
(a) $N_i(v_z,z)$ and (b) $N_i(v_{y'},z)$.
In Panel (b), $v_{y'}$ is evaluated in the reference frame, moving with $\vec{U}$.
}
\label{fig:phase}
\end{figure}

Theoretically, the particle motion in such a magnetic field
is characterized
by the curvature parameter, $\kappa=(R_{\rm min}/\rho_{\rm max})^{1/2}$,
where $R_{\rm min}$ is the minimum curvature radius of the magnetic field line
and $\rho_{\rm max}$ is the maximum Larmor radius of the particle.\citep{BZ86,BZ89}
This is best evaluated in the reference frame,
in which the electric field is transformed away.\citep{speiser65} 
Around the region of our interest,
the parabolic field model (Eq.~\ref{eq:B-fit})
gives $R_{\rm min}= 0.616$. 
The ion velocity leads to the maximum Larmor radius
$\rho_{\rm max} = (m_i/eB_z) |\vec{v}_i-\vec{U}|$
in the reference frame.
These two give the $\kappa$ parameter for individual ions in the reference frame.

\begin{figure*}[htbp]
\centering
\includegraphics[width={2\columnwidth},clip]{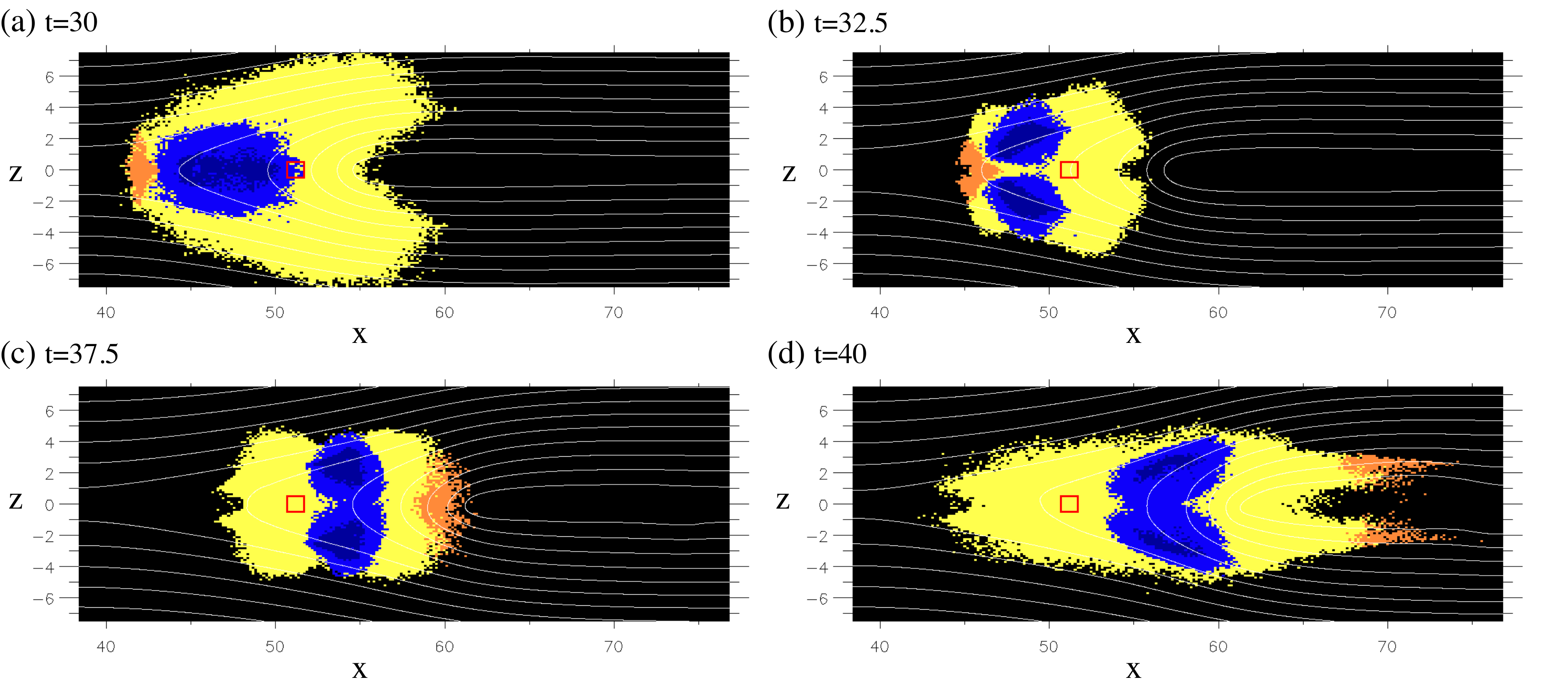}
\caption{(Color online)
Spatial distribution of the selected ions at (a) t=30, (b) t=32.5, (c) t=37.5, and (b) t=40.
The red box indicates the domain that we sample the ions.
The color corresponds the position in the VDF in Figure \ref{fig:outPDF}(a').
\label{fig:dist}
}
\end{figure*}
\begin{figure*}[htbp]
\centering
\includegraphics[width={2\columnwidth},clip]{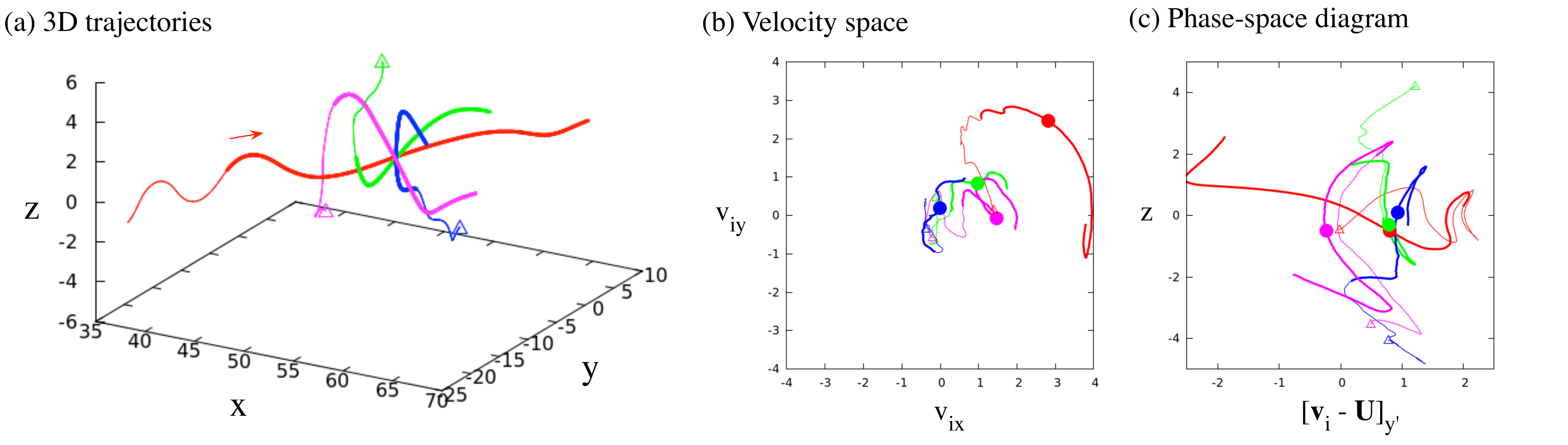}
\caption{
Reconstructed ion trajectories
(a) in the 3D $x$-$y$-$z$ space,
(b) in the $v_{x}$-$v_{y}$ velocity space, and
(c) in the $v_{y'}$-$z$ phase space.
The triangle marks the starting point at $t=15$.
The thick line indicates the trajectory during $30<t<40$.
In Panels (b) and (c), the circle marks the position at $t=35$.
In Panel (c), $v_{y'}$ is evaluated in the reference frame, moving with $\vec{U}$.
\label{fig:traj}
}
\end{figure*}

In the VDF in the $v_{x}$-$v_{y}$ space (Fig.~\ref{fig:outPDF}(a)),
we classify the ions into the following three groups:
(1) the global Speiser ions,
(2) the local Speiser ions, and
(3) the trapped ions.
The first population, the global Speiser ions are found
in the energetic tail, as indicated by the white arrow
in Figure \ref{fig:outPDF}(a).
A representative orbit is shown in red color in Figure \ref{fig:traj}.
This is a typical Speiser orbit.\citep{speiser65}
In order to distinguish them
from another Speiser ions in the next paragraph,
hereafter we call them the global Speiser ions. 
Their spatial distributions are shown in orange in Figure \ref{fig:dist}.
They are leaving the X-type region at $t=30$,
move on in the $x$ direction, and then
they are far away from the red box at $t=40$. 
Since the reconnection rate remains quasisteady after $t \gtrsim 20$,\citep{zeni11d}
these ions are continuously accelerated by the reconnection electric field $E_y$. 
Figure \ref{fig:outPDF}(d) shows the ion VDF in $v_{x}$-$v_{y}$ around the X-line,
integrated over $x, z \in [37.6,38.6] \times [-0.5,0.5]$ at $t=35$.
A similar energetic tail is found in the $+v_y$ direction in the VDF.
It represents the meandering ions in the $+y$ direction. 
As we depart from the X-line to the outflow direction,
the tail gradually changes its direction clockwise, and then
we see it in $x'$ in Figure \ref{fig:outPDF}(a). 
It is reasonable to see them in the ${x'}$ direction from $\vec{U}$
(the blue arrow in Fig.~\ref{fig:outPDF}(a)),
because the Speiser ions eventually escape along the magnetic field lines
outside the current sheet. 
The tail further rotates clockwise to the $v_x$ direction,
as seen in the ion orbit in Figure \ref{fig:traj}(b).
This is because the magnetic field lines outside the current sheet change
their direction to the $x$ direction in the farther downstream. } 
Near the X-line, the highest-energy ions have the typical speed of $v_{iy} \sim 3 c_{Ai}$
(Fig.~\ref{fig:outPDF}(d)).
Assuming $(E_y/c_{Ai}B_0) \sim (B_z/B_0) \sim 0.1$, one can estimate the energy gain,
$\Delta {\mathcal E} \sim e (0.1 c_{Ai}B_0) (3c_{Ai}/\Omega_{ci}) (B_0/B_z) \sim 3m_ic^2_{Ai}$. 
This is consistent with the typical velocity of the energetic ions
in Figure \ref{fig:outPDF}(a), $\sim 4 c_{Ai}$.

The second population, the local Speiser ions, is rounded
by the the white circle in Figure \ref{fig:outPDF}(a).
They are the main population. 
As shown in yellow in Figures \ref{fig:dist}(a) and \ref{fig:dist}(b),
they are in the upstream regions in the previous stages.
As reconnection evolves,
they travel backward along the field-lines to the midplane.\citep{drake09} 
The ions are locally reflected by $B_z$ around the midplane
through the Speiser orbit,\citep{speiser65} 
and then travels in the outflow direction.
The blue and green trajectories in Figure \ref{fig:traj} are relevant.
Both of them travels backward, and then
they are rotating to the $+x'$ direction while bouncing in $z$ around the midplane.
Basically, the Speiser motion is a combination of
a bounce motion in $z$ and a half gyration by $B_z$
in the reference frame with no electric field. 
The half gyration by $B_z$ corresponds to
a half-ring surrounding the reference-frame velocity $\vec{U}$
in the velocity space. 
Since the magnetic field is in $\pm{x'}$ outside the midplane,
these ions are found in the upper-left side of the circle
in Figure \ref{fig:outPDF}(a). 
Such half-ring VDFs have been reported
in previous works with hybrid simulations.\citep{nakamura98,lot98,arz01}
In Figure \ref{fig:outPDF}(a),
our VDF looks like a semicircle rather than a half-ring,
because the initial upstream temperature is high. 
The typical curvature parameters for local Speiser ions are $\kappa = 0.2$--$0.3$. 
There are $3$--$4$ bulges in the toroidal direction in Figure \ref{fig:outPDF}(a).
They are attributed to the midplane crossings.
Theoretically, the ions cross the midplane
$[(\sqrt{2}\kappa)^{-1}+1] \approx 3.3$--$4.5$ times
during the half rotation about $B_z$.\citep{chen86,chen92}
Here, we applied the relation $\hat{H}=1/(2\kappa^{4})$
to Eq.~(9) in Ref.~\onlinecite{chen86}.
In Figure \ref{fig:traj},
both blue and green orbits cross the midplane twice.
This is attributed to the limited duration of the run, $0\le t \le 40~[\Omega^{-1}_{ci}]$,
while it takes $(\pi/0.17) \sim 18.5$ to turn around $B_z$.
In Figure~\ref{fig:traj}(b),
the blue and green orbits exhibit a similar rotation of different phases in the velocity space.
Assuming that they travel in the similar path,
one can extrapolate the path to the 0th and 3rd crossings
to see $2$--$4$ midplane-crossings.
This needs to be verified by a larger PIC simulation in future.
They are mapped around $0.5<[\vec{v}-\vec{U}]_{y'}<1$
in the phase-space diagram (Figs.~\ref{fig:phase}(b) and \ref{fig:traj}(c)).

The third population, the trapped ions, is found in the small spot
near the reference-frame velocity $\vec{U}$ (the red cross) in Figure \ref{fig:outPDF}(a).
They enter the exhaust from the inflow regions near the X-type region, and then
they continue to bounce in $z$ during the system evolution. 
In Figure \ref{fig:dist}, they are shown in blue.
They are near the midplane at $t=30$ (Fig.~\ref{fig:dist}(a)),
temporally depart from the midplane in $\pm z$ (Fig.~\ref{fig:dist}(b)),
move back to the red box at $t=35$, and then
keep bouncing in $\pm z$ (Fig.~\ref{fig:dist}(c)).
The ions are finally located at $x \sim 55$--$59$ at $t=40$ (Fig.~\ref{fig:dist}(d)),
although this stage seems to be influenced by the periodic boundary effects.
In general, these ions keep bouncing across the ion current layer in $|z| \lesssim 3$
in the $\vec{U}$-moving frame.
They are trapped around the ion current layer in the reference frame.
This differs from the magnetic mirror confinement in the $\kappa \gg 1$ regime,
because the corresponding curvature parameter is $\kappa=0.3$--$0.4$. 
A typical orbit is shown in magenta in Figure \ref{fig:traj}(a).
Unlike other orbits, it does not substantially move in the $y$ direction.
The trajectory is characteristic in the $v_{y'}$-$z$ space (Fig.~\ref{fig:traj}(c)). 
The ion travels in an arc-shaped path in this phase space,
except for the last stage which is influenced by the periodicity.
The trapped ions are indeed evident
in the phase-space diagram in Figure \ref{fig:phase}(b). 
They are confined along a narrow path from $([\vec{v}-\vec{U}]_{y'},z)=(-0.2,0)$ to $(0.4,2.3)$,
They travels oppositely from the Speiser ions in the $y'$ direction
$[\vec{v}-\vec{U}]_{y'}<0$ at the midplane.

In the reference frame,
the trapped ions travel through regular orbits, discussed by \citet{chen86}.
To understand the orbits,
we further carry out supplemental test-particle simulations in a parabolic field. 
We will describe the setup in the Appendix and
we only present key results here. 
Figure \ref{fig:poincare} shows Poincar\'{e} section of surface
in the normalized velocity space $\dot{x}$-$\dot{y}$ at $z=0$
for a representative parameter $\kappa=0.36178$. 
This corresponds to the ion VDF in $v_{x'}$-$v_{y'}$ at the midplane
at a certain energy in the reference frame.
The regular orbits are found in the onion-ring region in the bottom half.
They are separated from an outer stochastic region by a boundary,
called a Kolmogorov-Arnold-Moser (KAM) surface.
The fixed-point $(\dot{x}$,$\dot{y}) \approx (0, -0.67)$
corresponds to an ``8''-shaped stationary orbit,
as shown in Figure \ref{fig:orbit}. 
A weak perturbation leads to an oscillation around the 8-shaped orbit
(for example, see Fig.~5 in Ref.~\onlinecite{chen92}),
projected to a ring in the Poincar\'{e} map.
In the real space, the magenta orbit in Figure \ref{fig:traj}(a)
is a combination of an ``8''-like regular motion
in the rotated coordinate system in the $\vec{U}$-moving frame. 
From test-particle orbits in the parabolic field,
we reconstruct the ion VDF and the $v_{y'}$-$z$ diagram
in Figure \ref{fig:reconstruct}. 
Features in the PIC data are excellently reproduced,
such as a small spot in the VDF 
and the narrow path in the phase-space diagram
(Figs.~\ref{fig:reconstruct}(c,d)). 
The VDF actually consists of ions with various $\kappa$ values.
The regular orbits are always found
near the fixed-point at $(0, -0.68) \sim (0, -0.65)$ in the Poincar\'{e} map
in the $\kappa \lesssim 0.4$ regime.\citep{wang94} 
In Figure \ref{fig:phase}(b),
it is surprising to see such a clear path in the phase-space diagram 
after a few bounces. 
We attribute this to the nature of the chaos system. 
In the outer stochastic region,
neighboring particles tend to diverge at a finite time.\citep{chen92}
These ions are quickly scattered, and then
the regular-orbit ions remain unscattered.

These regular orbits are modulated by the electric fields.
As seen in Figure \ref{fig:vertical}(d),
there exist a weak $E_x$ and
a strong $E^*_z$ at the separatrices ($z \sim 2.5$)
in the $\vec{U}$-moving frame. 
These polarization electric fields surround
an electrostatic potential in the exhaust.\citep{kari07} 
It is known that
this $E_x$ slightly kicks the ions in the $x$ direction.\citep{aunai11a}
Careful inspection of Figure \ref{fig:reconstruct}(d)
reveals minor differences between
the test-particle orbits and the ion phase-space diagram.
Even though the particle energies are set similar,
the former ranges $|z|<3$, while the latter is confined in $|z|<2.5$.
The ${y'}$-velocities are shifted by $\Delta v=0.2$--$0.3$. 
This is because the separatrix electric field $E^*_z$ ($z \sim 2.5$)
reflects the ions into the exhaust. 
The phase-space path are shifted in $+v_{iy'}$
in order to maintain the trapping condition $\oint [\vec{v}_i-\vec{U}]_{y'}dt \approx 0$.
Since the separatices move away in ${\pm}z$ in the further downstream,
such modulation will be found only near the X-type region.


To be fair,
there are another minor population around $(v_x,v_y)\sim(-0.5,5.5)$
outside the velocity space in Figure \ref{fig:outPDF}(a).
Their density is very low, corresponding to the blue region
($<13$ ions per the cell) in Figure \ref{fig:outPDF}(a).
Their total number is less than $0.1\%$ of the entire particle number in the same volume.
We find that they originally come from the initial preexisting current sheet far downstream
and they are being reflected by the piled-up magnetic field $B_z$. 
Since they do not contribute the dynamics, and
since we do not expect them in much larger systems,
they are excluded from Figure \ref{fig:outPDF}(b). 

Figures \ref{fig:outPDF}(e) and \ref{fig:outPDF}(f) are
the electron VDFs in the same volume,
$x, z \in [50.7,51.7] \times [-0.5,0.5]$ at $t=35$. 
The average velocity $\vec{\bar{V}}_e = (1.33,-0.37,0)$ and
the reference-frame velocity $\vec{U}$ are indicated by
the blue and red crosses in Figure \ref{fig:outPDF}(e), respectively.
The electrons are isotropic in $v_x$-$v_y$ and oval-shaped in $v_x$-$v_z$.
This is reasonable because they are gyrating about $B_z$.
The electron temperature is substantially high and
its thermal velocity is $5$. 
The curvature parameter is $\kappa=0.9$--$1.6$ for typical electrons.
Interestingly, as seen in the contour lines,
the VDF is somewhat flat-top shaped, rather than Maxwellian. 
All these results indicate that
the electrons are mostly gyrotropic ($\kappa\gtrsim 1$).

\section{Discussion}
\label{sec:discussion}

We have discussed that the electron shock separates
the upstream kinetic region and the downstream Hall-physics region. 
We do not know what controls its position from the X-line.
The previous works reported that
the jet extends several 10s of $d_i$,\citep{kari07,shay07}
but many factors could control the jet extent in $x$.
For example, we observe a short electron jet in our preliminary $m_i/m_e=400$ run,
suggesting that electron physics controls the length. 
The plasma temperature in the inflow region can be another factor,
because it increases the pressure in a reconnected flux tube ahead of the election jet.
In addition, recent works suggest that
the electron jet exists in a limited parameter range.\citep{goldman11,le13}
Despite all these uncertainties,
the electron shock is located in the middle of the outflow region in this work,
and so
we can separate the ion current layer from the upstream electron kinetic region.

An important question is whether
the violation of the ion ideal condition leads to
magnetic dissipation or magnetic diffusion,
which could control the reconnection process. 
In Sec.~\ref{sec:results},
we already mentioned that there is no energy dissipation
in the ion current layer. 
Here we examine the magnetic diffusion,
going back to its original meaning.
Let $\vec{R}$ be the nonideal part of the Ohm's law
with respect to a velocity field $\vec{V}$,
\begin{eqnarray}
\vec{E} + \vec{V} \times \vec{B}
= \vec{R}
\label{eq:ohm}
.
\end{eqnarray}
The nonideal term modifies the induction equation,
\begin{eqnarray}
\frac{\partial}{\partial t} \vec{B}- \rot (\vec{V}\times\vec{B})
+ \rot \vec{R}
= 0.
\label{eq:induction}
\end{eqnarray}
The magnetic flux penetrating a comoving surface is preserved,
when the flux~preservation condition is met,\citep{newcomb58,stern66}
\begin{eqnarray}
\rot \vec{R} = 0.
\label{eq:fluxpreservation}
\end{eqnarray}
This is often referred to as the ``frozen-in'' condition. 
A constant resistivity $\vec{R}=\eta \vec{j}$ leads to
a diffusion term $\rot \vec{R} = -(\eta/\mu_0)\Delta \vec{B}$
in the induction equation (Eq. \ref{eq:induction}).
This is the origin of the term ``magnetic diffusion.''
Such magnetic diffusion is a special case of the violation of the flux preservation.
Importantly,
these notions of nonidealness, frozen-in, and magnetic diffusion
depend on one's choice of the velocity field $\vec{V}$. 
Therefore it is possible that $\vec{R}(\vec{V}_i) \ne \vec{R}(\vec{V}_e)$,
as often found in a kinetic plasma. 
Keeping this in mind,
we expect that the ``ion diffusion region'' involves
a diffusion-like violation of the flux preservation,
with respect to the ion velocity $\vec{V}_i$.

At the midplane, considering the symmetry in $z$ and $\partial_y=0$, 
we obtain $\rot \vec{R} \approx (0, 0, \partial_x R_y)$.
Regarding electrons,
the ideal condition is fairly satisfied in $y$ in the shock-downstream,
${R}_y(\vec{V}_e) \approx 0$.
Consequently, we obtain
$\rot \vec{R}(\vec{V}_e) \approx 0$.
Regarding ions,
we have found that
${R}_y(\vec{V}_i) = [\vec{E}+\vec{V}_i\times\vec{B}]_y$ is almost flat in $x$
in Sec.~\ref{sec:overview} (the red arrow in Fig.~\ref{fig:horizontal}(b)).
This leads to $\rot \vec{R}(\vec{V}_i) \approx 0$.
Since the MHD velocity $\vec{V}_{\rm MHD}$ is obtained by
linear interpolation of $\vec{V}_{i}$ and $\vec{V}_{e}$,
we further obtain $\rot \vec{R}(\vec{V}_{\rm MHD}) \approx 0$.
The magnetic flux is preserved with respect to
$\vec{V}_i$, $\vec{V}_e$, and $\vec{V}_{\rm MHD}$.
In other words, magnetic flux is frozen into all these flows. 
Since no magnetic diffusion takes place in the ion, electron, or MHD flows,
the ion current layer is not a part of any diffusion regions.

The ion VDFs give us further insights into
the macroscopic properties in the ion current layer. 
Here we limit our discussion near the midplane
and so we implicitly assume $\vec{U}=\vec{w}$.
In Sec.~\ref{sec:overview}, we have pointed out that
the ion outflow velocity $V_{ix}\sim 0.7$ is
substantially slower than
the inflow Alfv\'{e}n speed $c_{A,in} \approx 1.62$,
while theories expect that they are comparable. 
In fact, the ion outflow is usually sub-Alfv\'{e}nic
near the reconnection site 
in recent PIC simulations.\citep{kari07,drake09,penny11,liu12}
The local Speiser distribution is responsible for this discrepancy.
In Figure \ref{fig:outPDF}(a),
the half-ring or semicircle VDF of the local Speiser ions
tells us that
the ion velocity $\vec{\bar{V}}_i$ does not meet $\vec{U}$.
The relative velocity $(\vec{\bar{V}}_i - \vec{U})$
arises from the meandering motion in $y'$. 
In the original coordinate system,
the ion outflow speed $V_{ix}$ is slower than
the \textbf{E} $\times$ \textbf{B} speed $w_{x}$ by
$|\vec{V}_i-\vec{U}| \sin \alpha$. 
The reference-frame speed $U_x=1.2$ is still slower than,
but better agrees with the inflow Alfv\'{e}n speed of $c_{A,in} \approx 1.62$.
Therefore, the sub-Alfv\'{e}nic outflow is partially an apparent effect. 
We have also found that the ion outflow speed
counter-intuitively decreases from $V_{ix}=0.8$ to $0.7$ in the shock-downstream region. 
This is because the local-Speiser ions start to join the midplane
and then they slows down the average speed $V_{ix}$. 
%
The violation of the ion ideal condition
is a logical consequence of the local Speiser motion.
Since $\vec{B} \approx B_z\vec{e}_z$,
the fact $\vec{\bar{V}}_{i} \ne \vec{U} = \vec{w}$ immediately yields
$\vec{E}+\vec{\bar{V}}_i\times\vec{B} \ne 0$ around the midplane. 
The popular ion-scale proxy of the reconnection site,
$[\vec{E} + \vec{V}_i\times\vec{B}]_y \ne 0$, is just an oblique projection. 

Possible signatures of the trapped ions were found
in earlier hybrid simulations,\citep{lot98,arz01}
although the authors explored other important issues. 
For example,
we recognize small spots near the Speiser-ring
in Figure 6d in Ref.~\onlinecite{lot98} and in Figure 3-(2) in Ref.~\onlinecite{arz01}. 
These VDFs were measured in the central current layer
in the outflow exhaust in consistent with our results. 
The role of the trapped ions was also discussed
in a thin current sheet (TCS) model in the magnetotail.
\citep{zelenyi02,malova10}
It was pointed out that
the trapped ions cannot be major population,
because they carry opposite currents near the midplane ($[\vec{v}_i-\vec{U}]_{y'}<0$)
to distort the current sheet structure.\citep{zelenyi02} 
In our case, their number is $\lesssim 10\%$ of the total number of ions in the volume,
and so they are unlikely to modify the current sheet. 
The trapped ions were expected to appear
as a stipe in the velocity space (in Fig. 3(b) in Ref.~\onlinecite{malova10}).
This is qualitatively similar to our VDF (Fig.~\ref{fig:outPDF}(a)). 
In the present work, we have found the trapped ions
for the first time in a self-consistent PIC simulation.
Examining their path in the real, velocity, and phase spaces,
we have identified that they are confined in regular orbits. 
The trapped ions are not major contributors to the macroscopic properties
around the midplane, due to their low density.
Judging from their orbits, they are in the $-v_{y'}$ side of
the reference-frame velocity $\vec{U}$ in the $v_x$-$v_y$ space.
Thus they do not modify $[\vec{E}+\vec{V}_i\times\vec{B}]_{y'}$.
Since the electric current is in $y'$,
they will not lead to the energy dissipation,
$D_e \approx j_{y'} [\vec{E}+\vec{V}_{i}\times\vec{B}]_{y'} \approx 0$.

Similar discussion can be applied to electrons
in the super-Alfv\'{e}nic electron jet
in the shock-upstream (Fig.~\ref{fig:outflow}(b)).
At the midplane, the magnetic field-lines are so bent that
the curvature radius is less than electron Larmor radius. 
Our inspection gives $\kappa = 0.2$-$0.3$ for typical electrons around $(x,z) \approx (43.2, 0)$.
The electron VDF should consist of local Speiser electrons. 
Since electrons gyrate oppositely from ions,
the Speiser semicircle has to be on the opposite side than
in Figure \ref{fig:outPDF}(a).
This explains
why the electron bulk speed $V_{ex}$ outruns the \textbf{E} $\times$ \textbf{B} speed $w_x$
and
why the electron ideal condition is violated 
$[\vec{E}+\vec{V}_e\times\vec{B}]_y \approx (w_{x}-V_{ex}) B_z < 0$ in the jet.\citep{kari07,shay07} 
Analyzing the electron Ohm's law,
\citet{hesse08} found that the electron nonidealness stems from a diamagnetic effect. 
Since the meandering motion during the Speiser motion is of diamagnetic-type,
their argument holds true. 
We do not see trapped electrons,
probably because electrons are easily scattered by waves,
but this is left for future investigation. 
Regardless of ions or electrons,
particles exhibit nongyrotropic motions
in a thin current sheet in the $\kappa \ll 1$ regime.
Since they do not fully gyrate about $B_z$,
the ideal conditions can be violated.

Throughout the paper,
we have assumed $\partial_x \approx 0$ in our discussion.
However, the system actually has an $x$-dependence. 
This is important for high-energy particles
whose Larmor radii are comparable to the scale length in $x$.
Burkhart et al.\cite{bkt91b,chaos92} studied
particle chaos in an X-type configuration,
$\vec{B}(x,z)= B_0 (z/L) \vec{e}_x + B_n (x/\lambda) \vec{e}_z$ and $\vec{E}=0$. 
They observed two KAM surfaces on each sides of the X-line
in the Poincar\'{e} map (e.g., Fig.~1a in Ref.~\onlinecite{bkt91b}). 
These regular orbits are similar to ones in our parabolic case. 
In fact, these ions mainly bounce in $z$ and move a little in $x$,
and so the $x$-variation does not make a big difference.
The authors further found that
the chaos is sensitive to the magnetic aspect ratio $b_n=B_n/B_0=L/\lambda$
around X-type region. 
They observed the regular orbits when $b_n \lesssim 0.3$.\cite{chaos92}
Judging from the separatrix slopes, our results correspond to $b_n=0.22$.
So far, the X-type chaos model does not contradict our results.
It is necessary to extend the chaos model to make a better comparison.



At this stage of investigation,
we do not know the large-scale picture of the outflow exhaust. 
In our run, the ions have crossed the midplane only a few times.
The trapped ions will remain around the midplane,
as long as the ion current layer is stable. 
If unstable, the trapped ions will be scattered away,
and so they will be observed only near the X-type region.
Farther downstream,
we may eventually see an MHD outflow.
In such a case, unlike the electron shock,
we will see a gradual transition from the ion current layer to the MHD flow, 
because the ion outflow speed is slower than
the \textbf{E} $\times$ \textbf{B} speed, $V_{ix}<w_{x}\approx V_{ex}$. 

NASA's upcoming Magnetospheric Multiscale (MMS) mission will observe
reconnection sites in the Earth's magnetosphere.
It will measure the ion VDF with an energy resolution of 20\% and
an angular resolution of $11.25 \times 11.25$ degrees. 
This will be sufficient to identify our VDFs. 
In addition, it is useful to estimate the \textbf{E} $\times$ \textbf{B} velocity.
MMS will measure the electric field with an accuracy of 0.5 [mV/m] in the spin plane.
When $B_z = 1$ [nT], we can estimate $\vec{w}$ with an accuracy of 500 [km/s]. 
In our case, the Speiser-ring radius in the ion VDF is
comparable with the inflow Alfv\'{e}n speed.
Given that it is ${\sim}2000$ [km/s] in the magnetotail,
MMS will tell the relation between $\vec{w}$ and the Speiser-ring. 
The electron moment data may improve the estimate,
because $\vec{w} \approx \vec{V}_e$ in the ion current layer.

\section{Summary}
\label{sec:summary}

We have investigated 
kinetic aspects of the ion current layer
at the center of the reconnection outflow exhaust,
by means of a PIC simulation and supplemental test-particle simulations. 
Regarding the ion fluid properties,
the ion current layer features
the sub-Alfv\'{e}nic outflow speed and
the violation of the ion ideal condition. 
Since the nonidealness does not involve magnetic dissipation nor magnetic diffusion,
the ion current layer does not appear to be
a key region to control the reconnection process. 
Regarding the kinetic physics, we have found that
the ion VDFs consist of the following three populations,
the global Speiser ions from the X-line,
the local Speiser ions from the separatrices,
and the trapped ions bouncing around the midplane. 
These motions are understood in the reference frame
by rotating and shifting the coordinate system.
The trapped ions are the first demonstration of the regular orbits,
originally discussed by \citet{chen86},
in a self-consistent PIC simulation. 
The ion fluid properties are intuitively explained by these particle motions.
The sub-Alfv\'{e}nic outflow speed is partially attributed to
an oblique projection of the local Speiser ions. 
This leads to the ion nonidealness in the out-of-plane direction
$\vec{E} + \vec{V}_i\times\vec{B} \ne 0$.
These VDFs would be observable with the upcoming MMS spacecrafts
near reconnection sites.

\begin{acknowledgments}
One of the authors (SZ) appreciates discussions with H. Yoshida.
We thank two anonymous reviewers for their careful evaluation of this manuscript.
This research was supported
by Grant-in-Aid for Young Scientists (B) (Grant No. 25871054)
and
by NINS Program for Cross-Disciplinary Study.
\end{acknowledgments}

\appendix

\section{Particle dynamics in a 1D parabolic field}

Here we summarize the approach of \citet{chen86}
in a parabolic field,
$\vec{B}=B_0(z/L)\vec{e}_x+B_n\vec{e}_z$ and $\vec{E}=0$. 
We consider the nonrelativistic motion,
\begin{eqnarray}
m\frac{d\vec{v}}{dt} = e(\vec{v}\times\vec{B}).
\end{eqnarray}
After an appropriate normalization,\citep{BZ86,wang94}
one obtains the following nonlinear system characterized by $\kappa$,
\begin{eqnarray}
\ddot{x} &=& \kappa \dot{y} \\
\ddot{y} &=& z\dot{z}-\kappa \dot{x} \\
\ddot{z} &=& -z\dot{y}
.
\end{eqnarray}
We normalize the Hamiltonian to $h=\frac{1}{2}(\dot{x}^2+\dot{y}^2+\dot{z}^2)=\frac{1}{2}$,
while \citet{chen86} employed a different normalization, $\hat{H}=1/(2\kappa^{4})$. 

\begin{figure}[htbp]
\centering
\includegraphics[width={\columnwidth},clip]{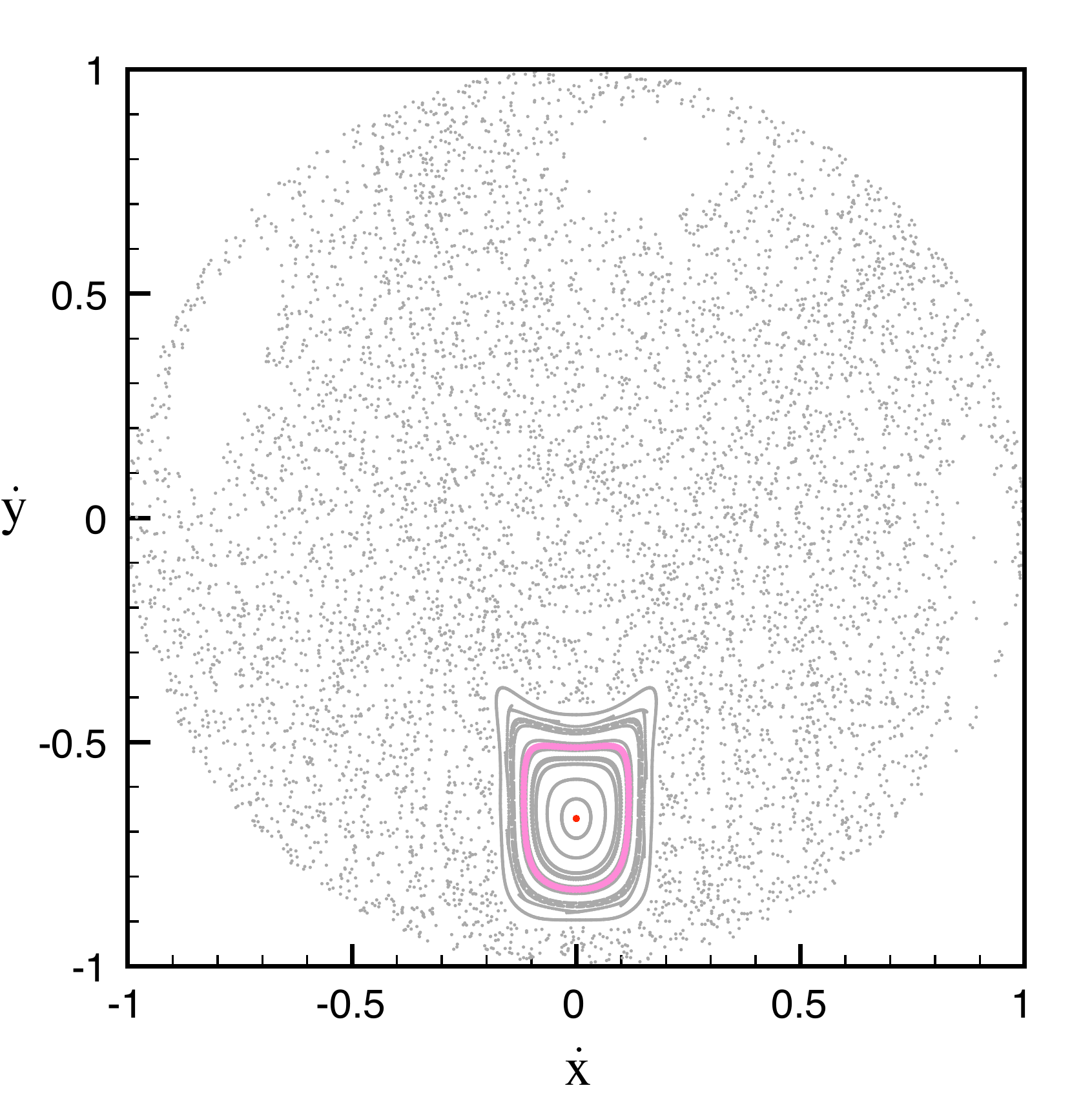}
\caption{(Color online)
Poincar\'{e} map for $\kappa=0.36178$ in the $\dot{x}$-$\dot{y}$ space.
The fixed point corresponds to the stationary orbit.
The closed circle in magenta corresponds to a regular orbit with weak perturbation.
\label{fig:poincare}
}
\end{figure}

\begin{figure}[htbp]
\centering
\includegraphics[width={\columnwidth},clip]{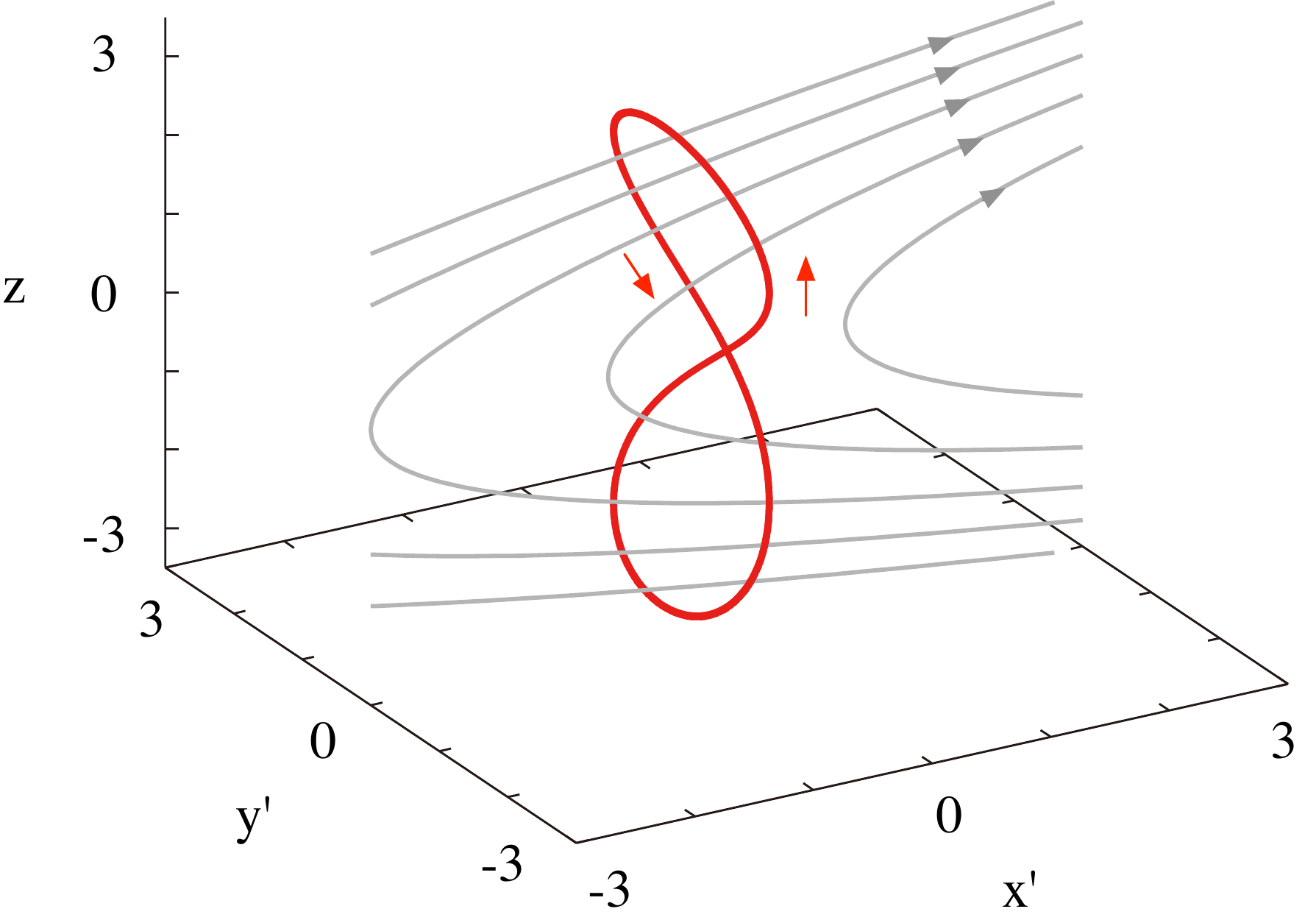}
\caption{(Color online)
Stationary orbit for $\kappa=0.36178$,
rescaled in the simulation units in the $x'$-$y'$-$z$ coordinate system
in the reference frame.
\label{fig:orbit}
}
\end{figure}

Using test particle simulations,
we evaluate $\dot{x}$ and $\dot{y}$ at the cross section of the midplane ($z=0$)
to visualize the type of particle orbits in a Poincar\'{e} map.
One can easily translate it to a popular $x$-$\dot{x}$ map
through the canonical momentum conservation,
$\dot{y}=\frac{1}{2}{z}^2-\kappa x$.
Figure \ref{fig:poincare} shows the Poincar\'{e} map for $\kappa=0.36178$.
Inside the onion-ring region,
a fixed-point $(\dot{x},\dot{y})=(0,-0.67046)$ corresponds to a stationary regular orbit.
Traveling through an ``8''-shaped orbit, the particle hits the same place
in the $\dot{x}$-$\dot{y}$ space at each midplane crossing. 
Figure \ref{fig:orbit} shows the trajectory in the units of the PIC simulation. 
In Figure \ref{fig:poincare},
the closed circles around the fixed-point correspond to
the regular orbits with weak perturbation.
The particles remain on the same circles at their midplane crossings. 
Outside the circles, the less-structured regions correspond to stochastic orbits,
whose orbits are difficult to predict.
Three low-density cavities correspond to the transient Speiser orbits.

\begin{figure}[htbp]
\centering
\includegraphics[width={\columnwidth},clip]{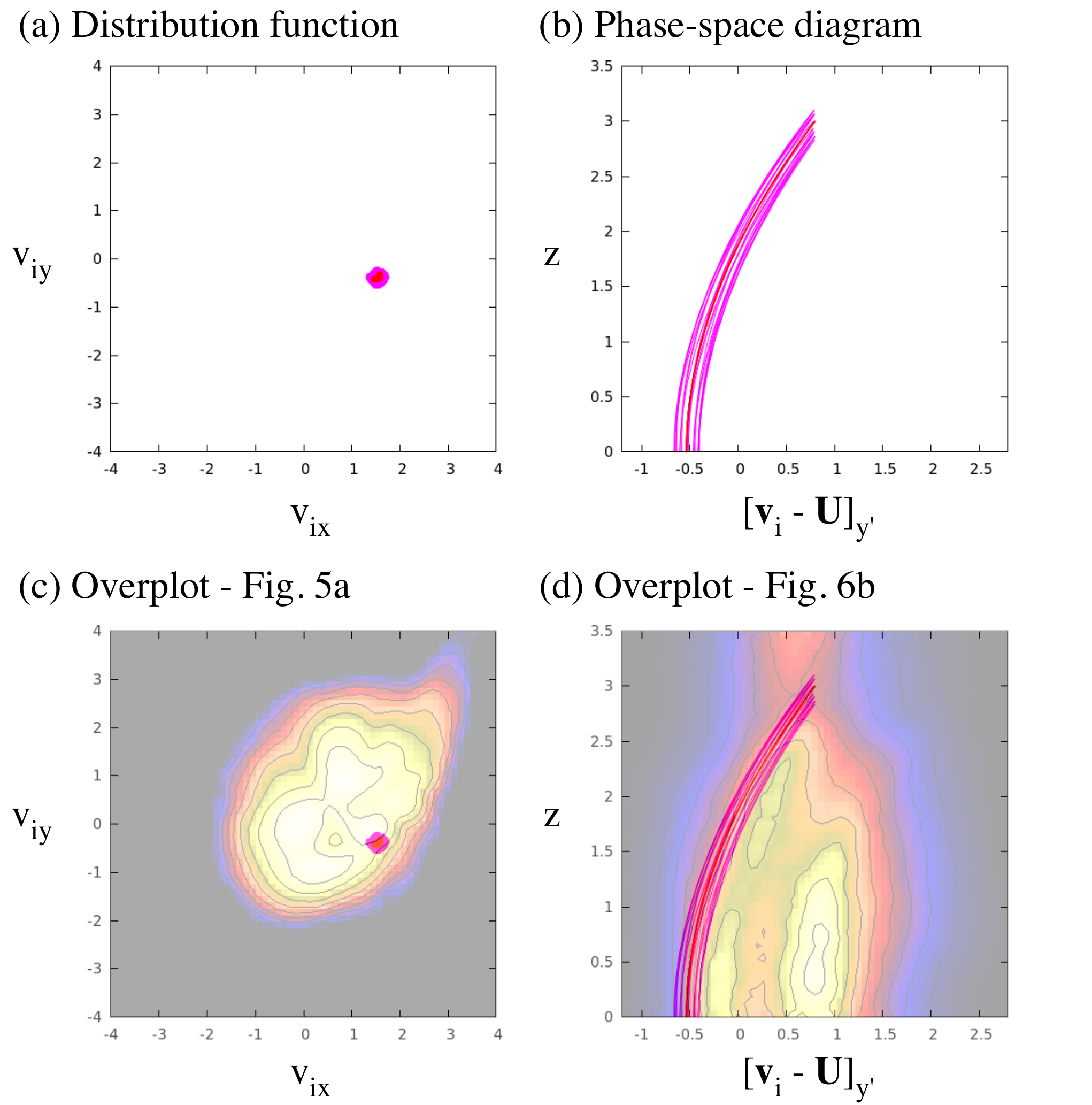}
\caption{(Color online)
Ion properties, predicted by the orbit theory.
The two regular orbits in Figure \ref{fig:poincare} are used.
(a) The velocity distribution function in $v_{x}$-$v_{y}$ and
(b) the phase-space diagram in $v_{y'}$-$z$ in the reference frame.
The bottom panels compare them with Figures \ref{fig:outPDF}(a) and \ref{fig:phase}(b).
\label{fig:reconstruct}
}
\end{figure}

We reconstruct several ion properties from the orbits.
In Figure \ref{fig:reconstruct},
the two regular orbits are projected to
the PIC simulation coordinate system
in the $v_{x}$-$v_{y}$ and $v_{y'}$-$z$ spaces.
The red and magenta orbits are
the stationary one and the weakly-perturbed one
in Figure \ref{fig:poincare}.
Figure \ref{fig:reconstruct}(a) is consistent with
the small separate spot in the ion VDF (Fig.~\ref{fig:outPDF}(a)).
Figure \ref{fig:reconstruct}(b) explains
the narrowly stretched path in the phase-space diagram
(Fig.~\ref{fig:phase}(b)). 

As far as we have investigated,
the fixed-point is located at $(\dot{x},\dot{y})=(0,-0.65) \sim (0,-0.68)$ and
the regular orbits occupy a similar domain
when $\kappa \lesssim 0.4$. 
The domain disappears when $\kappa > 0.53$.\citep{wang94}
In the $\kappa \rightarrow 0$ limit,
the configuration is asymptotic to
a current sheet with antiparallel fields.
Considering zero drift in an exact solution,
one obtains the stationary condition,\citep{parker57,sonnerup71}
\begin{eqnarray}
k = [(1/2) (1-v_{n}/v_{\perp})]^{1/2}=0.9092,
\end{eqnarray}
where $k$ is the orbit parameter, $v_n$ is the canonical momentum $v_y$ at the midplane,
and $v_{\perp}=(v_y^2+v_z^2)^{1/2}$ is the perpendicular speed.
From $v_{n}/v_{\perp}=v_y/|v|$,
we obtain an asymptotic fixed-point $(\dot{x},\dot{y})=(0,-0.653)$.

\end{document}